# A DUAL CHARACTERIZATION OF SELF-GENERATION AND EXPONENTIAL FORWARD PERFORMANCES

By Gordan Žitković[1]

*University of Texas at Austin*

We propose a mathematical framework for the study of a family of random fields—called *forward performances*—which arise as numerical representation of certain rational preference relations in mathematical finance. Their spatial structure corresponds to that of utility functions, while the temporal one reflects a Nisio-type semigroup property, referred to as self-generation. In the setting of semimartingale financial markets, we provide a dual formulation of self-generation in addition to the original one, and show equivalence between the two, thus giving a dual characterization of forward performances. Then we focus on random fields with an exponential structure and provide necessary and sufficient conditions for self-generation in that case. Finally, we illustrate our methods in financial markets driven by Itô-processes, where we obtain an explicit parametrization of all exponential forward performances.

**1. Introduction.** The present paper aims to contribute to the fruitful and successful literature on utility maximization and optimal investment in stochastic financial markets. Born in the seminal work of Merton [24, 25], the theory has been further developed by Pliska [33], Cox and Huang [6], Karatzas et al. [19], He and Pearson [16], Kramkov and Schachermayer [22], Cvitanić, Schachermayer and Wang [7], Karatzas and Žitković [21] and many others. In the setting similar to the one employed in here—namely, incomplete semimartingale markets with utility functions defined on the whole real line—the pertinent contributions include those of Frittelli [14], Bellini and Frittelli [4], Schachermayer [36], Owen and Žitković [32] and others.

Received November 2008; revised March 2009.

[1]Supported in part by the NSF under award number DMS-07-06947. Any opinions, findings and conclusions or recommendations expressed in this material are those of the author and do not necessarily reflect those of the National Science Foundation.

The notion of *forward performance* or *forward utility* has appeared in the literature recently, and in various forms, in the work of Choulli, Henderson, Hobson, Li, Musiela, Stricker and Zariphopoulou (see [5, 17, 26, 27, 28, 29, 30]). It refers to a family of interrelated state-dependent utility functions parametrized by the positive time axis $[0, \infty)$. The glue holding these utility functions together is the following economic principle of consistency: a rational economic agent should be indifferent between two random pay-offs as long as one can be produced from the other using a costless dynamic trading strategy in a financial market. We lay no claim to any originality in its formulation. In fact, it has existed in various forms in the financial literature for a long time. Recently, it has been used in the context of risk measures and their generalizations (see [13] and [15], among many other instances). An axiomatic treatment of a class of forward performances by Zariphopoulou and Žitković in [38] is based on an implenetation of this idea in the context of the risk-measure theory, but without a fixed finite investment horizon.

The main goal of the present manuscript is to establish a solid mathematical footing for the notion of forward performances, provide a dual characterization and illustrate the obtained results. Mathematically, the economic consistency criterion described above translates into a Nisio-type semigroup property which we call self-generation. The obstacles in the analysis, construction and characterization of self-generating random fields come from several directions. First, the level of generality needed for financial applications usually surpasses that of a finite-state-variable (i.e., finite-dimensional Markov setting) and deals with random fields of utilities whose dependence structure is quite general. Therefore, the classical PDE-based control-theoretic tools no longer apply. Second, the market models we consider are typically incomplete, as the complete case degenerates in a certain sense, and lacks interesting mathematical or economic content. Incompleteness or, in analytic language, lack of strict ellipticity renders the analysis much more delicate; in particular, as is well known in the utility maximization literature (see [7, 19, 21, 22] or [39] for a sample), the dual formulation introduces nontrivial functional-analytic difficulties. Our third obstacle is the lack of a terminal time-point. In fact, in the presence of such a point, say $T > 0$, there is a one-to-one correspondence between forward performances (*random fields*) and state-dependent utilities (*functions*) defined on $[0, T]$. The whole semigroup is then constructed via a backward projection-type operation, starting from its value at $T$. This situation is completely analogous to the one found in elementary martingale theory: martingales on the finite horizon $[0, T]$ come in a one-to-one correspondence with their terminal values. On the other hand, when no time horizon is specified, there is no obvious candidate for the terminal value, and the construction or characterization of self-generating random fields is far from trivial.

The present manuscript starts with a construction of a proper framework for the study of utility random fields in the context of financial markets driven by locally-bounded semimartingales. In this context, we define random fields dual in the convex-analytic sense to the utility random fields and study their properties. Our first result states that a utility random field is self-generating if and only if its dual is self-generating, where the notion of self-generation in the dual case is defined naturally over sets of probability measures (local-in-time local-martingale measures for the asset-price processes). The first benefit of the dual formulation is that it always admits an optimizer, i.e., the minimum in its definition is always attained, unlike in the case of the original, primal, problem where such a property is not required. This point is worth stressing as all the other treatments of forward performances, save the one in [38], explicitly require that the corresponding utility maximization problems admit maximizers, typically in a restricted domain. A removal of such a difficult-to-check requirement, as illustrated in the sequel, allows for much more flexibility in the theory and makes a symmetric dual characterization possible. In addition to its pleasant analytic properties, the dual formulation admits a convenient simplification when the utility random field takes some of the special forms often used in applications. The second focus of the present paper is the study of utility random fields of the exponential form. Here, the dual problem "separates" and we are able to use it to give a complete characterization of all self-generating exponential utility random fields. Finally, we restrict our attention to the case of continuous financial models based on Itô-process dynamics and describe explicitly all exponential forward performances in that setting. By using an argument based on the optional decomposition theorem, we find that the class of all forward utilities is essentially no larger than the class of examples presented heuristically in [29]. In particular, continuity of the market dynamics together with the self-generation requirement automatically implies the continuity of the utility random field.

With the notion of forward performances still being in its infancy, the literature on the subject is rather scarce. In addition to the work of Choulli, Henderson, Hobson, Li, Musiela, Stricker and Zariphopoulou mentioned above, the only other instance we are aware of is [3], where the authors focus on a notion of self-generation defined under much more stringent assumptions, such as market continuity and applicability of the Itô–Wentzel formula.

One of our major goals is generality, especially in the first part of the paper. That adds to the technical difficulty of the presentation and involves several novel results pertaining to the convex-duality analysis of random fields. In order not to interfere with the presentation flow for the reader only interested in the final product, those are relegated to the Appendix. The rest of the paper is presented in the logical order: the modeling environment is set up in Section 2. Section 3 introduces the notions of self-generation and the related dual concept and states the equivalence of the two. Section 4 deals with the utility random fields of the exponential type, while Section 5 studies the Itô-process models.

## 2. The financial set-up.

2.1. *The market model.* Let $(S^0; S) = (S_t^0, S_t^1, \ldots, S_t^d)_{t \in [0,\infty)}$ be a $(d+1)$-dimensional càdlàg semimartingale on a filtered probability space $(\Omega, \mathcal{F}, \mathbb{F}, \mathbb{P})$, where $\mathbb{F} = (\mathcal{F}_t)_{t \in [0,\infty)}$ satisfies the usual conditions of right-continuity and $\mathbb{P}$-completeness and $\mathcal{F}_0$ is trivial, i.e., generated by the $\mathbb{P}$-null sets. The $d$-dimensional vector $S$ models the price process of the $d$ risky assets, while $S^0$ corresponds to a risk-free asset. As usual, we quote all asset-prices in units of $S^0$. Operationally, this amounts to the simplifying assumption $S^0 \equiv 1$, which will hold throughout.

In order to render the presentation simpler and the theory standard, we assume that $S$ is locally bounded. Most of what follows can be extended to a more general setting in which $S$ admits unpredictable unbounded jumps, but at a cost of overwhelming additional technical complexity. The class of examples in Section 5 is presented in the setting of Itô-process models and the reader interested solely in those can assume from the outset that stock-prices follow Itô-processes.

2.2. *Admissible portfolios.* An $\mathbb{F}$-predictable process $\pi = (\pi_t)_{t \in [0,\infty)}$ is said to be an *admissible portfolio* (*process*) if:

1. $\pi$ is $S$-integrable on $[0, T]$, for each $T \geq 0$, in the sense of stochastic-integration theory for semimartingales (see [34]), and
2. for any $T \geq 0$, there exists a constant $a > 0$ (possibly depending on $\pi$ and $T$, but not on the state of the world) such that the *gains process* $X^\pi$, given by $X_t^\pi = \int_0^t \pi_u \, dS_u$, $t \geq 0$, is bounded from below by $-a$, for all $t \in [0, T]$, $\mathbb{P}$-a.s.

The set of all admissible portfolio processes is denoted by $\mathcal{A}$. A separate notation for the set of all portfolio processes giving rise to bounded gains processes will be quite useful below: we set $\mathcal{A}_{\text{bd}} = \mathcal{A} \cap (-\mathcal{A})$.

2.3. *No free lunch with vanishing risk on finite horizons.* The natural assumption of no arbitrage is routinely replaced in literature by a slightly stronger, but still economically feasible assumption of *no free lunch with vanishing risk* (NFLVR). In our case, we do not require NFLVR to hold on the entire time-horizon $[0, \infty)$—that would lead to too strong a restriction on the available class of models. Instead, we impose the local condition *no free lunch with vanishing risk on finite horizons* (NFLVRFH).

ASSUMPTION 2.1. For each $T \geq 0$, there exists a probability measure $\mathbb{Q}$, defined on $\mathcal{F}_T$, with the following properties:

1. $\mathbb{Q} \sim \mathbb{P}|_{\mathcal{F}_T}$, where $\mathbb{P}|_{\mathcal{F}_T}$ is the restriction of the probability measure $\mathbb{P}$ to $\mathcal{F}_T$, and
2. each component of $S$ is a $\mathbb{Q}$-local martingale on $[0, T]$.

The set of all measures $\mathbb{Q}$ with the above properties will be denoted by $\mathcal{M}_T^e$. When we loosen the requirement of equivalence in Assumption 2.1 to the one of absolute continuity, we get a possibly bigger set which we denote by $\mathcal{M}_T^a$. The measures in $\mathcal{M}_T^e$ ($\mathcal{M}_T^a$) will be called *finite-horizon equivalent* (*absolutely-continuous*) *local martingale measures* on $[0, T]$.

We leave it to the reader to check that Assumption 2.1 implies the following relation for all $0 \leq T_1 \leq T_2$

$$\mathcal{M}_{T_1}^e = \{\mathbb{Q}|_{\mathcal{F}_{T_1}} : \mathbb{Q} \in \mathcal{M}_{T_2}^e\}.$$

In other words, the restriction map turns the family $(\mathcal{M}_t^e)_{t \in [0,\infty)}$ into an inversely directed system:

(2.1) $$\{1\} \leftarrow \mathcal{M}_{T_1}^e \leftarrow \mathcal{M}_{T_2}^e \leftarrow \cdots.$$

In general, such a system will not have an inverse limit, i.e., there will exist no set $\mathcal{M}_\infty^e$ with the property that $\mathcal{M}_T^e = \{\mathbb{Q}|_{\mathcal{F}_T} : \mathbb{Q} \in \mathcal{M}_\infty^e\}$ for all $T \geq 0$. In other words, even though the market may admit no arbitrage (free lunch with vanishing risk) on any finite interval $[0, T]$, there might exist an arbitrage opportunity if we allow the trading horizon to be arbitrarily long. Therefore, we give the following definition.

DEFINITION 2.2. A market model $(S_t)_{t \in [0,\infty)}$ is said to be *closed* if there exists a set $\mathcal{M}_\infty^e$ of probability measures $\mathbb{Q}$ equivalent to $\mathbb{P}$ such that

$$\mathcal{M}_T^e = \{\mathbb{Q}|_{\mathcal{F}_T} : \mathbb{Q} \in \mathcal{M}_\infty^e\}.$$

REMARK 2.3. Most market models used in practice are not closed. The simplest example is the Samuelson's model, where the filtration is generated by a single Brownian motion $(B_t)_{t \in [0,\infty)}$, and the price of the risky asset satisfies $dS_t = S_t(\mu\, dt + \sigma\, dB_t)$, for some constants $\mu \in \mathbb{R}$, $\sigma > 0$. For $T \geq 0$, the only element in $\mathcal{M}_T^e$ corresponds to a Girsanov transformation which turns $B_s + \frac{\mu}{\sigma} s$, $s \in [0, T]$ into a Brownian motion. It is well known that in the limit as $T \to \infty$, these transformations become "more and more singular" with respect to $\mathbb{P}|_{\mathcal{F}_T}$, and no $\mathbb{Q}$ as in Definition 2.2 can be found (see [20], remark on page 193).

In fact, it is useful to think of the closed market models as essentially finite-horizon, perhaps under a time change. Moreover, just like classical notions of admissibility (boundedness from below, etc.) rule out "nonphysical" arbitrage opportunities in the form of doubling schemes, the requirement of NFLVRFH does not insist on closedness, but still rules out arbitrages based on strategies that have a predetermined deterministic upper bound on time duration.

It will be useful in the sequel to introduce the so-called *density processes* for local martingale measures: for $T \geq 0$ and $\mathbb{Q} \in \mathcal{M}_T^e$, the process $Z^{\mathbb{Q}} = \{Z_t^{\mathbb{Q}}\}_{t \in [0,T]}$ is defined as the càdlàg version of the conditional expectation $\mathbb{E}[\frac{d\mathbb{Q}}{d(\mathbb{P}|_{\mathcal{F}_T})}|\mathcal{F}_t]$, $t \in [0,T]$. In fact, the assumption of NFLVRFH guarantees that each $Z^{\mathbb{Q}}$ can be extended (nonuniquely) to a positive martingale $(Z_t)_{t \in [0,\infty)}$ on $[0,\infty)$, so that:

1. $Z$ is a strictly positive martingale with $Z_0 = 1$, and
2. $ZS$ is a (component-wise) local martingale.

The set of all such processes $Z$ will be denoted by $\mathcal{Z}^e$. If the requirement of strict positivity is replaced by the one of nonnegativity, the obtained, larger, family is denoted by $\mathcal{Z}^a$. The elements of $\mathcal{Z}^e(\mathcal{Z}^a)$ are called *positive* (*nonnegative*) *densities*. It can be argued that in our setting, they are a natural proxy for the family of sets of measures from Assumption 2.1. In fact, Assumption 2.1 is equivalent to the statement $\mathcal{Z}^e \neq \varnothing$. Furthermore, a financial market is closed if and only if $\mathcal{Z}^e$ contains a uniformly integrable element. We conclude the discussion of densities with a useful convention: we shall often use quotients of the form $Y_t/Y_s$, $s \leq t$, where $Y$ is a nonnegative càdlàg supermartingale (a density process, in particular), even when the random variable $Y_s$ takes the value 0 with positive probability. The supermartingale property and the regularity of paths of $Y$ imply that $Y_t = 0$, a.s., on $\{Y_s = 0\}$, which allows us to set $Y_t/Y_s := 1$ on $\{Y_s = 0\}$, so that:

1. $Y_s \frac{Y_t}{Y_s} = Y_t$, a.s., for all càdlàg supermartingales $Y$, and
2. $\mathbb{E}[\frac{Y_t}{Y_s}|\mathcal{F}_s] = 1$, a.s., when $Y$ is a nonnegative càdlàg martingale.

**3. Utility random fields, self-generation and a dual characterization.** Having described the financial environment in the previous section, we turn to a class of random fields used in behavioral modelling of economic agents.

3.1. *Utility random fields and their associated value fields.*

DEFINITION 3.1. A mapping $U : \Omega \times [0, \infty) \times \mathbb{R} \to \mathbb{R}$ is called a *random field* if it is measurable with respect to the product $\mathcal{O} \times \mathcal{B}(\mathbb{R})$ of the optional $\sigma$-algebra on $\Omega \times [0, \infty)$ and the Borel $\sigma$-algebra on $\mathbb{R}$. A *utility random field* is a random field such that the following three conditions hold:

1. *Utility conditions.* There exists $\Omega' \in \mathcal{F}$ such that $\mathbb{P}[\Omega'] = 1$ and for all $(\omega, t) \in \Omega' \times [0, \infty)$, the mapping $x \mapsto U(\omega, t; x)$ is:
   (a) a strictly concave, strictly increasing $C^1(\mathbb{R})$-function, and
   (b) satisfies the Inada conditions
   $$\lim_{x \to -\infty} \frac{\partial}{\partial x} U(\omega, t; x) = \infty, \qquad \lim_{x \to \infty} \frac{\partial}{\partial x} U(\omega, t; x) = 0.$$
2. *Path regularity.* There exists $\Omega' \in \mathcal{F}$ with $\mathbb{P}[\Omega'] = 1$ such that the function $t \mapsto U(\omega, t; x)$ is càdlàg on $[0, \infty)$ for all $(x, \omega) \in \mathbb{R} \times \Omega'$.
3. *Integrability.* For each $x \in \mathbb{R}$ and $T \in [0, \infty)$, $U(\cdot, T, x) \in \mathbb{L}^1(\mathcal{F}_T)$.

As usual in probability, we suppress the $\omega$ from the notation and write simply $U(t, x)$ in the sequel, unless we want to expressly stress the nondeterministicity of $U$.

REMARK 3.2. The reader should note that $U(t, x)$ is assumed to be finite-valued for all $x \in \mathbb{R}$. A parallel theory can be built for utility functions taking values in $[-\infty, \infty)$, i.e., in the case when $U(t, x)$ is only finite for $x \in (a, \infty)$ (or $x \in [a, \infty)$), for some $a \in \mathbb{R}$. As the authors have shown in [22], the duality theory in this case requires a lot of care and interesting but technical subtleties appear. Hence, we do not pursue it in the present paper.

In addition to natural requirements of Definition 3.1, we will usually impose the following, very mild technical condition which effectively precludes pathological appearance of noncountably-additive measures in the dual treatment. A theory without this requirement is possible, but, similarly to the case described in Remark 3.2, it will not be dealt with here as it would introduce a prohibitive amount of technicalities without any real benefit. Moreover, as we shall see in the proof of Theorem 4.4, this technical condition is automatically implied by the natural integrability conditions for the class of exponential utility random fields on which a large part of the present paper focuses.

DEFINITION 3.3. A utility random field $U$ is said to be *nonsingular* if for each $T \geq 0$, and for each nonincreasing sequence $\{D_n\}_{n \in \mathbb{N}}$ in $\mathcal{F}_T$ with $\bigcap_n D_n = \varnothing$, there exists a sequence $\{a_n\}_{n \in \mathbb{N}}$ in $(0, \infty)$ such that
$$a_n \to \infty \quad \text{and} \quad \limsup_n \frac{1}{a_n} \mathbb{E}[U(T, -a_n \mathbf{1}_{D_n})] \geq 0.$$

REMARK 3.4. The nonsingularity condition of Definition 3.3 is automatically satisfied for deterministic utility random fields $U(\omega, t; x) = U(t, x)$. Indeed, thanks to Inada conditions, we have $\lim_{x \to -\infty} -U(-x)/x = \infty$. So, for a sequence $\{D_n\}_{n\in\mathbb{N}}$ as above, we can find a sequence $\{a_n\}_{n\in\mathbb{N}}$ with $a_n \to \infty$ such that $-U(t,-a_n)\sqrt{\mathbb{P}[D_n]} \leq a_n$ for all $n \in \mathbb{N}$. Then

$$\limsup_n \frac{1}{a_n}\mathbb{E}[U(t,-a_n\mathbf{1}_{D_n})] = \limsup_n \left(\frac{U(t,-a_n)\mathbb{P}[D_n]}{a_n}\right) + \lim_n \frac{U(t,0)}{a_n} \geq 0.$$

More generally, one can apply the same argument to show that it is enough for the random field $U(T,x)$ to be $(x,\omega)$-uniformly bounded from below by a deterministic utility function. This can be further relaxed due to the fact that we are dealing with the expected value of $U$ in the statement of the condition.

For a $\sigma$-algebra $\mathcal{G} \subset \mathcal{F}$ and $I \subseteq [-\infty,\infty]$, let $\mathbb{L}^0(\mathcal{G};I)$ denote the set of all $\mathbb{P}$-a.s.-equivalence classes of $\mathcal{G}$-measurable (extended) random variables which take values in $I$, a.s. For $I = \mathbb{R}$, we simply write $\mathbb{L}^0(\mathcal{G})$.

The following definition introduces an object—called a *value field*—related to a utility random field $U$, which can be interpreted as the field of indirect utilities for an economic agent who invests in the financial market modeled by $S$ and uses $t$-slices of $U$ as utility functions. In order to make the analysis easier, we parametrize a value field by the initial and final time-points $t \leq T$ in the generic investment horizon $[t,T]$, as well as the initial (time-$t$) wealth $\xi$, which is allowed to be an $\mathcal{F}_t$-measurable random variable.

DEFINITION 3.5. Let $U$ be a utility random field. The *value field associated to* $U$ is a family of mappings $\{u(\cdot;t,T) : 0 \leq t \leq T < \infty\}$, with $u(\cdot;t,T) : \mathbb{L}^\infty(\mathcal{F}_t) \to \mathbb{L}^0(\mathcal{F}_t; \mathbb{R} \cup \{\infty\})$ given by

$$(3.1) \quad u(\xi;t,T) = \operatorname*{ess\,sup}_{\pi \in \mathcal{A}_{\mathrm{bd}}} \mathbb{E}\left[U\left(T, \xi + \int_t^T \pi_u \, dS_u\right) \Big| \mathcal{F}_t\right] \qquad \text{for } \xi \in \mathbb{L}^\infty(\mathcal{F}_t).$$

REMARK 3.6.

1. For $0 \leq t < T < \infty$, the integral $\int_t^T \pi_u \, dS_u$ should be interpreted as $\int_{t+}^T \pi_u \, dS_u$, i.e., the possible initial jump $\Delta S_t$ (where $S_{0-} = 0$) should be ignored.
2. Condition 4 of Definition 3.1 and the a.s.-monotonicity of the mapping $x \mapsto U(T,x)$ imply that $U(T,X) \in \mathbb{L}^1(\mathcal{F}_T)$, for any $X \in \mathbb{L}^\infty(\mathcal{F}_T)$. Therefore, $U(T, \xi + \int_t^T \pi_u \, dS_u) \in \mathbb{L}^1(\mathcal{F}_T)$ and, consequently, $u(\cdot;t,T)$ takes values in $(-\infty,\infty]$, a.s.

3.2. *Self-generation.* As already mentioned in the Introduction, self-generation is a mathematical expression of the replication-invariance property of a rational agent's preference structure when it admits a utility representation. Since the main focus of the present paper is on the mathematical analysis, we refrain from a deeper economic discussion of the concept. Instead, we direct the reader to [38] for a risk-measure-theoretic approach, or to the forthcoming in-depth discussion of the decision-theoretic and axiomatic foundations of the forward utilities and the notion of self-generation in [40]. Finally, we note that self-generation is related to a form of a Nisio-type semigroup property. The Nisio semigroup (introduced in [31]) is a successful attempt at expressing the Bellman's dynamic programming principle in analytic terms, typically as a semigroup of nonlinear operators. In our case, loosely speaking, the operators that form the semigroup are the maximization operators $U \mapsto \operatorname{ess\,sup}_{\pi \in \mathcal{A}_{\mathrm{bd}}} E[U(T, \cdot + \int_t^T \pi_u\, dS_u)|\mathcal{F}_t]$.

DEFINITION 3.7. We say that a utility random field $U$ is *self-generating* or a *forward performance* if $u(\xi; t, T) = U(t, \xi)$, a.s., for all $0 \le t \le T < \infty$ and all $\xi \in \mathbb{L}^\infty(\mathcal{F}_t)$, i.e., if

$$(3.2) \qquad U(t, \xi) = \operatorname*{ess\,sup}_{\pi \in \mathcal{A}_{\mathrm{bd}}} \mathbb{E}\bigg[U\bigg(T, \xi + \int_t^T \pi_u\, dS_u\bigg)\bigg|\mathcal{F}_t\bigg] \qquad \text{a.s.}$$

for $0 \le t \le T < \infty$ and all $\xi \in \mathbb{L}^\infty(\mathcal{F}_t)$.

REMARK 3.8. The important novel feature of our definition of self-generation—and this is where our notion differs from that in the work of Musiela and Zariphopoulou or Berrier et al.—is that we do not require that the essential supremum in (3.2) be attained. This variation opens the door to a more general analysis as one does not need to specify the exact domain (admissibility class) for the utility maximization problems. It is well known [especially in the case of utility functions defined over $(-\infty, \infty)$] that the precise choice of the said domain is a nontrivial matter and that it, in general, depends directly on the utility function used.

Let us also mention that the requirement that (3.2) hold for all $\xi \in \mathbb{L}^\infty(\mathcal{F}_t)$ can be shown to be equivalent to the seemingly weaker requirement where (3.2) is assumed to hold only for constant $\xi$. We choose this version to strengthen the characterization results below, and to keep in line with the structure of the results in Appendix A.

3.3. *Duality for utility random fields.* The use of convex duality in utility maximization (and optimal stochastic control in general) has proven extremely fruitful. As we shall see below, analysis of utility random fields is no exception. We start with a straightforward translation of the well-known Fenchel–Legendre conjugacy to the random-field case.

For a utility random field $U$, we define the *dual random field* $V : \Omega \times [0, \infty) \times (0, \infty) \to \mathbb{R}$, by

$$(3.3) \qquad V(t, y) = \sup_{x \in \mathbb{R}}(U(t, x) - xy) \qquad \text{for } t \ge 0, y \ge 0.$$

The value at $y = 0$, given by $V(t,0) = \sup_{x \in \mathbb{R}} U(t,x) \in (-\infty, \infty]$ can be adjoined to the definition of $V$, and we will use it in the sequel whenever the need arises without explicit mention. We do not include it in the definition above because of the fact that it can ruin the otherwise pleasant finite-valuedness which follows from the Inada conditions [Definition 3.1, 2(b)].

PROPOSITION 3.9. *The dual random filed $V$ given by (3.3) inherits the following properties from the utility random field $U$:*

1. *$V$ is measurable with respect to the product $\mathcal{O} \times \mathcal{B}(0,\infty)$ of the optional $\sigma$-algebra on $\Omega \times [0,\infty)$ and the Borel $\sigma$-algebra on $(0,\infty)$.*
2. *There exists $\Omega' \in \mathcal{F}$ with $\mathbb{P}[\Omega'] = 1$ such that for each $(\omega, t) \in \Omega' \times [0,\infty)$, the mapping $y \mapsto V(\omega, t; y)$, $y > 0$, is:*
    (a) *strictly convex, continuously differentiable, and*
    (b) *satisfies $\lim_{y \to 0} \frac{\partial}{\partial y} V(\omega, t; y) = \infty$, $\lim_{y \to \infty} \frac{\partial}{\partial y} V(\omega, t; y) = \infty$.*
3. *There exists an event $\Omega'$ with $\mathbb{P}[\Omega'] = 1$ such that for all $(y, \omega) \in [0, \infty) \times \Omega'$ the functions $t \mapsto V(\omega, t; y)$, $t \geq 0$, are right-continuous and admit no discontinuities of second-order.*
4. *For each $\zeta \in \mathbb{L}^0_+(\mathcal{F}_T)$, we have $V(T, \zeta) \geq U(T, 0)$. Moreover, $\max(0, -V(T, \zeta)) \in \mathbb{L}^1(\mathcal{F}_T)$.*

PROOF. The properties of the dual random field in Proposition 3.9 follow directly from the corresponding properties in Definition 3.1 of the (primal) utility random field $U$. The only part that needs comment is, perhaps, 3. It follows from the fact that pointwise convergence of a sequence of convex functions implies uniform convergence on compacts, as well as pointwise convergence of the corresponding convex conjugates (see Theorem 11.34, page 500 of [35]). □

The notion of the value random field transfers to the dual case. However, in this setting, the domain of optimization is chosen so that the full duality relationship can be derived.

DEFINITION 3.10. For $y > 0$ and $0 \leq t < T < \infty$, we define the *dual value field* $v(\cdot; t, T) : \mathbb{L}^0_+(\mathcal{F}_t) \to \mathbb{L}^0(\mathcal{F}_t; \mathbb{R} \cup \{\infty\})$,

$$(3.4) \qquad v(\eta; t, T) = \operatorname*{ess\,inf}_{\mathbb{Q} \in \mathcal{M}^a_T} \mathbb{E}[V(T, \eta Z^{\mathbb{Q}}_T / Z^{\mathbb{Q}}_t) | \mathcal{F}_t], \qquad \eta \in \mathbb{L}^0_+(\mathcal{F}_t).$$

In analogy with the notion of self-generation for the utility random fields, we introduce the same notion for the their duals.

DEFINITION 3.11. A dual utility random field $V$ is said to be *self generating* if $v(\eta; t, T) = V(t, \eta)$, i.e.,

$$(3.5) \qquad V(t, \eta) = \operatorname*{ess\,inf}_{\mathbb{Q} \in \mathcal{M}^a_T} \mathbb{E}[V(T, \eta Z^{\mathbb{Q}}_T / Z^{\mathbb{Q}}_t) | \mathcal{F}_t] \qquad \text{a.s.}$$

for all $0 \leq t < T < \infty$ and all $\eta \in \mathbb{L}^0_+(\mathcal{F}_t)$.

The main technical result, whose proof is quite lengthy and occupies most of Appendix A (Theorem A.5 and Corollary A.6), extends the conjugacy relationship from random fields to their value fields. The reader should note that almost no regularity conditions (except for the one of nonsingularity) are imposed. In particular, neither the primal nor the dual value field is assumed to be finite, or that the optimization problems in their definitions admit optimal solutions. In fact, it may very well happen that $u$ and $v$ have empty effective domains, i.e., that $u = v = \infty$ identically. A similar result, but for nonrandom utilities and under more stringent assumptions (finiteness of the dual value function and the existence of the dual optimizer in the class of equivalent martingale measures) has been proved in [37].

THEOREM 3.12. *Let $U$ be a utility random field satisfying the nonsingularity condition of Definition 3.3 and let $V$ be its dual random field as defined in (3.3). If $u$ and $v$ denote the primal and dual value fields, as defined in (3.1) and (3.4), then*

$$(3.6) \qquad u(\xi; t, T) = \operatorname*{ess\,inf}_{\eta \in \mathbb{L}^0_+(\mathcal{F}_t)} (v(\eta; t, T) + \xi \eta) \qquad a.s. \quad and$$

$$(3.7) \qquad v(\eta; t, T) = \operatorname*{ess\,sup}_{\xi \in \mathbb{L}^\infty(\mathcal{F}_t)} (u(\xi; t, T) - \xi \eta) \qquad a.s.$$

*for all $0 \leq t \leq T < \infty$, $\xi \in \mathbb{L}^\infty(\mathcal{F}_t)$ and $\eta \in \mathbb{L}^0(\mathcal{F}_t)$. Moreover, for each $\xi \in \mathbb{L}^\infty(\mathcal{F}_t)$, there exist $\hat{\eta} \in \mathbb{L}^0_+(\mathcal{F}_t)$, $\hat{\mathbb{Q}} \in \mathcal{M}^a_T$ such that*

$$u(\xi; t, T) = \mathbb{E}[V(T, \hat{\eta} Z_T^{\hat{\mathbb{Q}}}/Z_t^{\hat{\mathbb{Q}}})|\mathcal{F}_t] + \hat{\eta}\xi \in \mathbb{L}^0(\mathcal{F}_t, \mathbb{R} \cup \{\infty\}).$$

The following characterization follows directly from Theorem 3.12.

COROLLARY 3.13. *A nonsingular utility random field is self generating if and only if its dual random field is self generating.*

A more practical version of the characterization above, still in terms of the dual random field, is given in the following theorem. We adopt a definition of a submartingale slightly weaker than the standard one: a process $(Y_t)_{t \in [0, \infty)}$ is called a *submartingale* if $\min(Y_T, 0) \in \mathbb{L}^1(\mathcal{F}_T)$ and $\mathbb{E}[Y_T|\mathcal{F}_t] \geq Y_t$, a.s., for all $0 \leq t < T < \infty$, where we use an extended, $(-\infty, \infty]$-valued, version of the conditional expectation.

THEOREM 3.14. *Let $U$ be a nonsingular utility random field, and let $V$, given by (3.3), be its conjugate. Then the following two statements are equivalent.*

1. *U is self generating.*
2. *For each $y > 0$, we have:*
    (a) *the process $(V(t, yZ_t))_{t \in [0, \infty)}$ is a càdlàg submartingale for all $Z \in \mathcal{Z}^a$, and*
    (b) *there exists $Z \in \mathcal{Z}^a$ such that $(V(t, yZ_t))_{t \in [0, \infty)}$ is a martingale.*

*In particular, if the market is complete, i.e., if $\mathcal{Z}^e = \{Z\}$, then $U$ is self-generating if and only if the process $(V(t, yZ_t))_{t \in [0, \infty)}$ is a martingale for each $y > 0$.*

PROOF. We start by assuming that $U$ is self generating. By Corollary 3.13, the relation (3.5) holds. Therefore, for $Z \in \mathcal{Z}^a$, $y > 0$ and $0 \leq t \leq T < \infty$, we can simply pick $\eta = yZ_t \in \mathbb{L}^1_+(\mathcal{F}_t)$ and use (3.5) to conclude that $(V(t, yZ_t))_{t \in [0, \infty)}$ is a càdlàg submartingale. To show (b), we take $t = 0$, and fix an arbitrary $T > 0$. According to Theorem 3.12, for each $x \in \mathbb{R}$ there exists $\hat{\mathbb{Q}}(x) \in \mathcal{M}^a_T$ and $y(x) \geq 0$ such that

$$U(0, x) = \mathbb{E}[V(T, y(x) Z^{\hat{\mathbb{Q}}(x)}_T)] + xy(x).$$

Since $U(0, x) = \inf_{y > 0}(V(0, y) + xy)$ and $V(t, y(x) Z^{\hat{\mathbb{Q}}(x)}_t) + xy(x)$ is a submartingale on $[0, T]$, the following two conclusions must hold:

- $V(t, y(x) Z^{\hat{\mathbb{Q}}(x)}_t)$ is a martingale on $[0, T]$, and
- $U(0, x) = V(0, y(x)) + xy(x)$.

The conjugacy relationship between $U$ and $V$ forces the relationship $y(x) = \frac{\partial}{\partial x} U(0, x)$. Inada conditions imply that the mapping $x \mapsto y(x) = \frac{\partial}{\partial x} U(0, x)$ is onto $(0, \infty)$. Therefore, for each $y > 0$, there exists $x \in \mathbb{R}$ such that $y = y(x)$ and $V(t, yZ^{\hat{\mathbb{Q}}(x)}_t)$ is a martingale on $[0, T]$. The extension to $[0, \infty)$ follows by a simple "patch-up" over larger an larger time horizons $[0, T]$.

We start the proof of the converse implication $2 \Rightarrow 1$ by noting that 2 yields

$$V(t, \eta Z_t) \leq \mathbb{E}[V(T, \eta Z_T) | \mathcal{F}_t] \quad \text{a.s. for } Z \in \mathcal{Z}^a$$

as soon as $\eta = \sum_{k=1}^n y_k \mathbf{1}_{A_k}$ is a simple, positive and $\mathcal{F}_t$-measurable random variable. For a general $\eta \in \mathbb{L}^0_+(\mathcal{F}_t)$, let the sequence $\{\eta_n\}_{n \in \mathbb{N}}$ of simple functions in $\mathbb{L}^0_+(\mathcal{F}_t)$ be given by

$$\eta_n = \begin{cases} \lfloor \eta \rfloor_{1/n}, & \eta Z_T > \frac{\partial}{\partial x} U(T, 0), \\ \eta, & \eta Z_T = \frac{\partial}{\partial x} U(T, 0), \\ \lceil \eta \rceil_{1/n}, & \eta Z_T < \frac{\partial}{\partial x} U(T, 0), \end{cases}$$

where, for $x \in \mathbb{R}$, $\alpha > 0$, $\lfloor x \rfloor_\alpha = \sup\{k\alpha : k \in \mathbb{Z}, k\alpha \leq x\}$ and $\lceil x \rceil_\alpha = \inf\{k\alpha : k \in \mathbb{Z}, k\alpha \geq x\}$. The fact that $V(T, \cdot)$ is decreasing on $(-\infty, \frac{\partial}{\partial x} U(T, 0))$ and increasing on $(\frac{\partial}{\partial x} U(T, 0), \infty)$ implies that the sequence $\{\eta_n\}_{n \in \mathbb{N}}$ of simple, $\mathcal{F}_T$-measurable random variables has the following two properties:

- $\eta_n \to \eta$, a.s.,
- $V(T, \eta_n Z_T) \nearrow V(T, \eta Z_T)$, a.s.

Recall that $V(T, x) \geq U(T, 0) \in \mathbb{L}^1(\mathcal{F}_T)$. Then the monotone convergence theorem implies that

$$\begin{aligned}
\mathbb{E}[V(T, \eta Z_T) | \mathcal{F}_t] &= \lim_n \mathbb{E}[V(T, \eta_n Z_T) | \mathcal{F}_t] \\
&\geq \limsup_n V(t, \eta_n Z_t) = V(t, \eta Z_t) \quad \text{a.s.}
\end{aligned} \tag{3.8}$$

In particular, we have

$$V(t, \eta) \leq \operatorname*{ess\,inf}_{Z \in \mathcal{Z}^a} \mathbb{E}[V(T, \eta Z_T / Z_t) | \mathcal{F}_t]. \tag{3.9}$$

The equality in (3.9) follows by a similar argument, where all the inequalities are turned into equalities by the choice of the element $Z \in \mathcal{Z}^a$ for which $V(t, yZ_t)$ is a martingale. Therefore, $V$ is self-generating, and by Corollary 3.13, so is $U$. $\square$

**4. Utility random fields of the exponential type.** Our next task is to specialize the structure of the utility random field and to use Theorem 3.14 to provide a workable characterization of self generation.

DEFINITION 4.1. A random field $U : \Omega \times [0, \infty) \times \mathbb{R} \to \mathbb{R}$ is said to be *of the exponential type* if there exist stochastic processes $(A_t)_{t \in [0, \infty)}$ and $(\gamma_t)_{t \in [0, \infty)}$ such that

$$U(t, x) = -e^{-\gamma_t x + A_t} \quad \text{for } t \geq 0, x \in \mathbb{R}. \tag{4.1}$$

The choice of the form in (4.1) can be explained by the success that the use of exponential utility has had in the mathematical-finance literature (we single out the seminal contribution of [8] among a myriad of other important papers). Furthermore, as one varies the coefficient-processes $\gamma$ and $A$, one gets a remarkably flexible family of preference structures. Finally, as we shall see shortly, the duality theory is especially generous in the exponential case; in particular, it admits a detailed characterization of the forward utilities of the exponential type.

4.1. *A necessary and sufficient condition for self-generation.* Our analysis starts with the notion of relative entropy tailored to the exponential random fields.

DEFINITION 4.2. Let $(\gamma_t)_{t\in[0,\infty)}$ and $(A_t)_{t\in[0,\infty)}$ be adapted stochastic processes with $\gamma_T > 0$, a.s. and $\mathbb{E}[\exp(A_T)] < \infty$ for all $T \geq 0$. For $0 \leq t \leq T < \infty$, the *relative conditional $(\gamma, A)$-entropy on $[t,T]$*, denoted by $H(\cdot; t, T)$, is a functional acting on probability measures $\mathbb{Q} \ll \mathbb{P}|_{\mathcal{F}_T}$ on $\mathcal{F}_T$ with values in $\mathbb{L}^0(\mathcal{F}_t, \mathbb{R} \cup \{\infty\})$, given by

$$(4.2) \qquad H(\mathbb{Q}; t, T) = \mathbb{E}\left[h\left(\frac{1}{\gamma_T} Z_T^{\mathbb{Q}} / Z_t^{\mathbb{Q}}\right) - Z_T^{\mathbb{Q}} / Z_t^{\mathbb{Q}} \frac{1}{\gamma_T} A_T \Big| \mathcal{F}_t\right],$$

where, $h(y) = y \log(y) - y$, $y \geq 0$.

REMARK 4.3. Inequality $h(y) + \exp(x) \geq xy$, valid for all $x \in \mathbb{R}$, $y \geq 0$, and assumption $\mathbb{E}[\exp(A_T)] < \infty$, imply that $(h(Y) - YA_T)^- \in \mathbb{L}^1(\mathcal{F}_T)$, for all $Y \in \mathbb{L}^0(\mathcal{F}_T)$. Hence, $H$ takes values in $(-\infty, \infty]$.

While the processes $\gamma$ and $A$ are, initially, quite free in the specification of the class of exponential random fields, the following theorem shows that the requirement of self-generation places quite a significant restriction on their structure.

THEOREM 4.4. *Let the financial market be as in Section 2 and let $(\gamma_t)_{t\in[0,\infty)}$ and $(A_t)_{t\in[0,\infty)}$ be stochastic processes. Then for the exponential random field $U$, given by $U(t,x) = -e^{-\gamma_t x + A_t}$, $t \geq 0$, $x \in \mathbb{R}$, the following two statements are equivalent.*

1. *$U$ a self-generating utility random field.*
2. *The following three assertions hold:*
   (a) *$\gamma$ and $A$ are càdlàg semimartingales with $\gamma_T > 0$, for all $T \geq 0$, a.s., and $\mathbb{E}[\exp(A_T + n\gamma_T)] < \infty$, for all $T \geq 0$ and all $n \in \mathbb{N}$.*
   (b) *For all $0 \leq t < T < \infty$, and all $\mathbb{Q} \in \mathcal{M}_T^a$*

$$\mathbb{E}^{\mathbb{Q}}\left[\frac{1}{\gamma_T}\Big|\mathcal{F}_t\right] = \frac{1}{\gamma_t} \qquad \text{on } \{H(\mathbb{Q}; t, T) < \infty\}.$$

   (c) *For all $0 \leq t < T < \infty$,*

$$h\left(\frac{1}{\gamma_t}\right) - \frac{1}{\gamma_t} A_t = \operatorname*{ess\,inf}_{\mathbb{Q} \in \mathcal{M}_T^a} H(\mathbb{Q}; t, T) \qquad a.s.$$

*Furthermore, if $U$ is self-generating, it is automatically nonsingular.*

PROOF. $1 \Rightarrow 2$ We first show that a self-generating exponential random field $U$ must satisfy all three parts of statement 2.

2(a) *holds*. Definition 3.1 of the utility random field implies that $(\gamma_t x - A_t)_{t \in [0,\infty)}$ is an adapted and càdlàg process for each $x$. The constant process $\pi \equiv 0$ is in $\mathcal{A}_{\mathrm{bd}}$, so the self-generation property (3.2) and part 2 of Definition 3.1 imply that $(U(t,x))_{t \in [0,\infty)}$ is a càdlàg supermartingale, for each $x \in \mathbb{R}$. Therefore, its $C^2$-transformation $(\gamma_t x - A_t)_{t \in [0,\infty)}$ is a semimartingale, and so are both $\gamma$ and $A$. Finally, $\gamma_T > 0$, a.s. for all $T \geq 0$ by part 1 of Definition 3.1 and the random variable $\exp(A_T + n\gamma_T)$ is in $\mathbb{L}^1$ for each $n \in \mathbb{N}$ by part 3 of the same definition.

*$U$ is nonsingular.* Let $0 \leq T < \infty$ be arbitrary but fixed, and let $\{D_n\}_{n \in \mathbb{N}}$ be a decreasing sequence in $\mathcal{F}_T$ with $\bigcap_n D_n = \varnothing$. Since $U(T,m) \in \mathbb{L}^1(\mathcal{F}_T)$, for all $m \in \mathbb{N}$, we have

$$\lim_{n \to \infty} \mathbb{E}[\exp(m\gamma_T \mathbf{1}_{D_n} + A_T)] \to C \qquad \text{where } C = \mathbb{E}[\exp(A_T)] \in (0, \infty).$$

In particular, for each $m \in \mathbb{N}$, there exists $n_m \in \mathbb{N}$ such that

$$\mathbb{E}[\exp(m\gamma_T \mathbf{1}_{D_n} + A_T)] \leq 2C \qquad \text{for all } n \geq n_m.$$

We can choose the sequence $\{n_m\}_{m \in \mathbb{N}}$ to be strictly increasing so that the sequence $\{a_n\}_{n \in \mathbb{N}}$, defined by

$$a_n = \sup\{m \in \mathbb{N} : n_m \leq n\} \qquad \text{where } \sup \varnothing := 1,$$

takes values in $\mathbb{N}$ and converges to $\infty$ as $n \to \infty$. Then, since $n_{a_n} \leq n$ for large enough $n \in \mathbb{N}$, we have

$$0 \leq -\limsup_{n \to \infty} \frac{1}{a_n} \mathbb{E}[U(T, -a_n \mathbf{1}_{D_n})] = \liminf_{n \to \infty} \frac{1}{a_n} \mathbb{E}[\exp(a_n \gamma_T \mathbf{1}_{D_n} + A_T)]$$
$$\leq \limsup_{n \to \infty} \frac{2C}{a_n} = 0.$$

Thus, the condition of Definition 3.3 is fulfilled.

2(b) *and* 2(c) *hold*. An elementary calculation shows that the random field $V$, dual to $U$ in the sense of (3.3), has the form

(4.3) $$V(t, y) = h\left(\frac{1}{\gamma_t} y\right) - \frac{1}{\gamma_t} y A_t \qquad \text{for } t \geq 0 \text{ and } y \geq 0,$$

with the function $h$ as in Definition 4.2. Using the nonsingularity of $U$ established above, Corollary 3.13 implies that $V$ is self-generating, i.e., that

$$h\left(\frac{1}{\gamma_t} \eta\right) - \frac{1}{\gamma_t} \eta A_t = \operatorname*{ess\,inf}_{Z \in \mathcal{Z}^a} \mathbb{E}\left[h\left(\frac{1}{\gamma_t} \eta Z_T / Z_t\right) - \frac{1}{\gamma_t} \eta Z_T / Z_t A_T \Big| \mathcal{F}_t\right] \qquad \text{a.s.}$$

for each $\eta \in \mathbb{L}^0_+(\mathcal{F}_t)$. A rearrangement of terms yields that

$$
(4.4) \quad \begin{aligned}
&(\eta + h(\eta))\frac{1}{\gamma_t} + \eta\left(h\left(\frac{1}{\gamma_t}\right) - \frac{1}{\gamma_t}A_t\right) \\
&= \underset{\mathbb{Q}\in\mathcal{M}^a_T}{\operatorname{ess\,inf}}\left((\eta + h(\eta))\mathbb{E}^{\mathbb{Q}}\left[\frac{1}{\gamma_T}\Big|\mathcal{F}_t\right] + \eta H(\mathbb{Q};t,T)\right) \quad \text{a.s.}
\end{aligned}
$$

for all $0 \leq t \leq T < \infty$ and all $\eta \in \mathbb{L}^1_+(\mathcal{F}_t)$. In particular, if we set $\eta = \exp(z)$, for some $z \in \mathbb{R}$, and divide the inequality (4.4) throughout by $\exp(z) > 0$, we get

$$z\frac{1}{\gamma_t} + h\left(\frac{1}{\gamma_t}\right) - \frac{1}{\gamma_t}A_t \leq z\mathbb{E}^{\mathbb{Q}}\left[\frac{1}{\gamma_T}\Big|\mathcal{F}_t\right] + H(\mathbb{Q};t,T) \quad \text{a.s.}$$

for all $z \in \mathbb{R}$ and all $\mathbb{Q} \in \mathcal{M}^a_T$. Since both sides of the above inequality are linear functions (in $z$), we must have

$$\frac{1}{\gamma_t} = \mathbb{E}^{\mathbb{Q}}\left[\frac{1}{\gamma_T}\Big|\mathcal{F}_t\right] \quad \text{a.s. on } \{H(\mathbb{Q};t,T) < \infty\} \text{ for all } \mathbb{Q} \in \mathcal{M}^a_T,$$

and

$$(4.5) \quad h\left(\frac{1}{\gamma_t}\right) - \frac{1}{\gamma_t}A_t \leq \underset{\mathbb{Q}\in\mathcal{M}^a_T}{\operatorname{ess\,inf}} H(\mathbb{Q};t,T) \quad \text{a.s.}$$

The equality in (4.4) implies that the a.s.-equality holds in (4.5).

$2 \Rightarrow 1$ Let us assume that $U$ is a random field of the form (4.1) which satisfies 2(a), 2(b) and 2(c). We first check that the requirements of Definition 3.1 hold. Parts 1 and 3 follow directly from the càdlàg semimartingale property of $A$ and $\gamma$. Part 2 is a consequence of the elementary properties of the exponential function and the strict positivity of $\gamma$. Finally, part 4 follows from the requirement that $\exp(A_T + n\gamma_T) \in \mathbb{L}^1(\mathcal{F}_T)$ and the monotonicity of the mapping $x \mapsto \exp(\gamma_T x + A_T)$.

Next, the nonsingularity of $U$ follows as in the first part of the proof by condition 2(a) We can now use Theorem 3.14 to show self-generation. Indeed, the conditions 2(b) and 2(c) imply that the equation (4.4)—which is equivalent to self generation of the dual random field $V$—holds. □

4.2. *On condition 2(b) of Theorem 4.4.* Condition 2(b) of Theorem 4.4 immediately hints at replicability of the process $(1/\gamma_t)_{t\in[0,\infty)}$. This is, indeed, true either under a mild additional assumption on the market model, or when restricted to a certain, maximal, event. The purpose of this subsection is to expand on those assumptions. Our main conclusion is Proposition 4.7, which is preceded by two lemmas.

In addition to the existing notation, we introduce the following subset of $\mathcal{M}^a_T$:

$$\mathcal{M}^H_T = \{\mathbb{Q} \in \mathcal{M}^a_T : H(\mathbb{Q};0,T) < \infty\}.$$

For a probability measure $\mathbb{Q}$ on $\mathcal{F}_T$, we define the "support" $\operatorname{equ}_{\mathbb{P}}\mathbb{Q}$ of $\mathbb{Q}$ with respect to $\mathbb{P}|_{\mathcal{F}_T}$ and the aggregation $\mathcal{S}_T^H$ of all such supports over the class of finite-entropy martingale measures:

$$\operatorname{equ}_{\mathbb{P}}\mathbb{Q} = \left\{\frac{d\mathbb{Q}}{d(\mathbb{P}|_{\mathcal{F}_T})} > 0\right\} \in \mathcal{F}_T \quad \text{and} \quad \mathcal{S}_T^H = \{\operatorname{equ}_{\mathbb{P}}\mathbb{Q} : \mathbb{Q} \in \mathcal{M}_T^H\},$$

where any two sets whose symmetric difference is $\mathbb{P}$-null are identified.

LEMMA 4.5. *For $T \geq 0$, assume that $\mathcal{M}_T^H \neq \varnothing$. Then there exists a $\mathbb{P}$-a.s.-unique event $C \in \mathcal{F}_T$ such that:*

1. *for all $A \in \mathcal{S}_T^H$, $A \subseteq C$, a.s., and*
2. *$C = \operatorname{equ}_{\mathbb{P}}\tilde{\mathbb{Q}}$, a.s., for some $\tilde{\mathbb{Q}} \in \mathcal{M}_T^H$.*

PROOF. Let $\{A_n\}_{n\in\mathbb{N}}$ be a sequence in $\mathcal{S}_T^H$ with the property that $\mathbb{P}[A_n] \to m$, where $m = \sup\{\mathbb{P}[A] : A \in \mathcal{S}_T^H\}$. Moreover, for $n \in \mathbb{N}$, let $\mathbb{Q}_n \in \mathcal{M}_T^H$ be such that $A_n = \operatorname{equ}_{\mathbb{P}}\mathbb{Q}_n$, and let $\alpha_n = 2^{-n}(H(\mathbb{Q}_n;0,T) + \mathbb{E}[\exp(A_T)] + 1)^{-1}$, so that $0 < \alpha_n \leq 2^{-n}$. Then the sequence $\{\tilde{\mathbb{Q}}_n\}_{n\in\mathbb{N}}$ of probability measures defined by

$$\tilde{\mathbb{Q}}_n = \frac{\sum_{k=1}^n \alpha_k \mathbb{Q}_k}{\sum_{k=1}^n \alpha_k}$$

converges in the total-variation norm and, consequently, weakly in $\sigma(\mathbb{L}^1(\mathcal{F}_T), \mathbb{L}^\infty(\mathcal{F}_T))$ when we identify measures with their Radon–Nikodym derivatives with respect to $\mathbb{P}$. We denote its limit by $\tilde{\mathbb{Q}}$. The functional $H(\cdot;0,T)$ is easily seen to be convex and $\sigma(\mathbb{L}^1(\mathcal{F}_T), \mathbb{L}^\infty(\mathcal{F}_T))$-lower semi-continuous. Thus,

$$H(\tilde{\mathbb{Q}};0,T) = H\left(\lim_n \tilde{\mathbb{Q}}_n; 0, T\right) \leq \liminf_n H(\tilde{\mathbb{Q}}_n; 0, T)$$

$$\leq \liminf_n \frac{\sum_{k=1}^n \alpha_k H(\mathbb{Q}_n; 0, T)}{\sum_{k=1}^n \alpha_k} \leq \frac{1}{\sum_{k=1}^\infty \alpha_k} < \infty.$$

Using the fact that $\mathcal{M}_T^a$ is convex and closed with respect convergence in total variation, we conclude that $\tilde{\mathbb{Q}} \in \mathcal{M}_T^H$. Moreover, $C := \operatorname{equ}_{\mathbb{P}}\tilde{\mathbb{Q}} = \bigcup_{n\in\mathbb{N}} \operatorname{equ}_{\mathbb{P}}\mathbb{Q}_n$, and so, $\mathbb{P}[C] = m$. It remains to show that $C$ is maximal in the sense of a.s.-inclusion, and not only with respect to its size. Let us assume that there exists $A = \operatorname{equ}_{\mathbb{P}}\mathbb{Q}' \in \mathcal{S}_T^H$ with $\mathbb{P}[A \setminus C] > 0$. Using the same ideas as above, we conclude that the probability measure $\bar{\mathbb{Q}}$, given by $\bar{\mathbb{Q}} = \frac{1}{2}\tilde{\mathbb{Q}} + \frac{1}{2}\mathbb{Q}'$, lies in $\mathcal{M}_T^H$ and has the property $\mathbb{P}[\operatorname{equ}_{\mathbb{P}}\bar{\mathbb{Q}}] = \mathbb{P}[C \cup A] > m$— a contradiction. $\square$

The set $C$, whose existence is guaranteed by Lemma 4.5, will be denoted by $\max \mathcal{S}_T^H$. If $\mathcal{M}_T^H = \varnothing$, we set $\max \mathcal{S}_T^H = \varnothing$.

LEMMA 4.6. *Suppose that for $T \geq 0$, we have*

$$(4.6) \qquad \mathbb{E}^{\mathbb{Q}}\left[\frac{1}{\gamma_T}\right] = \frac{1}{\gamma_0} \qquad \forall \mathbb{Q} \in \mathcal{M}_T^H.$$

*Then there exists $\pi \in \mathcal{A}$ such that*

$$(4.7) \qquad \frac{1}{\gamma_T} = \frac{1}{\gamma_0} + \int_0^T \pi_u \, dS_u \qquad \text{on } \max \mathcal{S}_T^H, \text{ a.s.}$$

PROOF. Let $\tilde{\mathbb{Q}}$ be the element of $\mathcal{M}_T^H$ such that $\text{equ}_{\mathbb{P}} \tilde{\mathbb{Q}} = \max \mathcal{S}_T^H$. We first show that $f \in \bar{\mathcal{C}}_T^{\tilde{\mathbb{Q}}}$, where $\bar{(\cdot)}^{\tilde{\mathbb{Q}}}$ denotes the closure in $\mathbb{L}^1(\tilde{\mathbb{Q}})$ while $\mathcal{C}_T = \{\int_0^T \pi_u \, dS_u : \pi \in \mathcal{A}_{\text{bd}}\} - \mathbb{L}_+^\infty(\mathcal{F}_T)$ and $f = \frac{1}{\gamma_T} - \frac{1}{\gamma_0}$. Suppose, to the contrary, that $f \notin \bar{\mathcal{C}}_T^{\tilde{\mathbb{Q}}}$. The Hahn–Banach separation theorem, applied for the duality between $\mathbb{L}^1(\tilde{\mathbb{Q}})$ and $\mathbb{L}^\infty(\mathcal{F}_T)$ guarantees the existence of an element $\chi \in \mathbb{L}^\infty(\mathcal{F}_T)$ such that $\mathbb{E}_{\tilde{\mathbb{Q}}}[\chi \zeta] \leq 0$ for $\zeta \in \bar{\mathcal{C}}_T^{\tilde{\mathbb{Q}}}$ and $\mathbb{E}_{\tilde{\mathbb{Q}}}[\chi f] > 0$. Since $-\mathbb{L}_+^\infty(\mathcal{F}_T) \subseteq \bar{\mathcal{C}}_T^{\tilde{\mathbb{Q}}}$, we have $\chi \in \mathbb{L}_+^\infty(\mathcal{F}_T) \setminus \{0\}$, and we can assume, without loss of generality, that $\mathbb{E}_{\tilde{\mathbb{Q}}}[\chi] = 1$. Therefore, the random variable $\zeta^* = \chi \frac{d\tilde{\mathbb{Q}}}{d(\mathbb{P}|_{\mathcal{F}_T})} \in \mathbb{L}_+^1(\mathcal{F}_T)$ satisfies $\mathbb{E}[\zeta^*] = 1$ and $\mathbb{E}[\zeta^* \zeta] \leq 0$, for all $\zeta \in \mathcal{C}_T$. So, $\mathbb{Q}_1 \in \mathcal{M}_T^a$ with $\frac{d\mathbb{Q}_1}{d(\mathbb{P}|_{\mathcal{F}_T})} = \zeta^*$. Moreover, we have

$$\begin{aligned} H(\mathbb{Q}_1; 0, T) &= \mathbb{E}\left[h\left(\frac{1}{\gamma_T}\zeta^*\right) - \zeta^* \frac{1}{\gamma_T} A_T\right] \\ &= \mathbb{E}\left[\chi\left(h\left(\frac{1}{\gamma_T}Z_T^{\tilde{\mathbb{Q}}}\right) - Z_T^{\tilde{\mathbb{Q}}} A_T\right)\right] + \mathbb{E}\left[\chi \log(\chi) \frac{1}{\gamma_T} Z_T^{\tilde{\mathbb{Q}}}\right] < \infty, \end{aligned}$$

where the finiteness is substantiated by $\tilde{\mathbb{Q}} \in \mathcal{M}_T^H$ and assumption (4.6). We deduce that $\mathbb{Q}_1 \in \mathcal{M}_T^H$, thus reaching a contradiction with the conjunction of the fact that $\mathbb{E}_{\mathbb{Q}_1}[f] > 0$ and the assumption (4.6).

The newly established fact that $f \in \bar{\mathcal{C}}_T^{\tilde{\mathbb{Q}}}$ implies that there exists a sequence $\{f_n\}_{n \in \mathbb{N}}$ in $\mathcal{C}_T$ such that $f_n \to f$ in $\mathbb{L}^1(\tilde{\mathbb{Q}})$. Note that each $f_n$ can be represented as

$$f_n = \int_0^T \pi_u^n \, dS_u - g_n$$

for some $\pi^n \in \mathcal{A}_{\text{bd}}$, $g_n \in \mathbb{L}_+^\infty$, and

$$0 = \mathbb{E}_{\tilde{\mathbb{Q}}}[f] = \lim_n \mathbb{E}_{\tilde{\mathbb{Q}}}[f_n] = \lim_n \mathbb{E}_{\tilde{\mathbb{Q}}}\left[\int_0^T \pi_u^n \, dS_u - g_n\right] = \lim_n \mathbb{E}_{\tilde{\mathbb{Q}}}[-g_n].$$

Thus, $g_n \to 0$ in $\mathbb{L}^1(\tilde{\mathbb{Q}})$. Consequently, we can safely take $g_n = 0$, for all $n \in \mathbb{N}$, without affecting the $\mathbb{L}^1(\tilde{\mathbb{Q}})$-convergence of $f_n$ to $f$. By Theorem 15.4.7 in [9], there exists $\pi \in \mathcal{A}$ such that $f = \int_0^T \pi_u \, dS_u$. □

The following proposition, which effectively explains the role of condition 2(b) of Theorem 4.4 in the majority of interesting cases, follows directly from Lemmas 4.5 and 4.6.

PROPOSITION 4.7. *Suppose that the exponential utility random field* $U(x) = -e^{-\gamma_t x + A_t}$ *is self-generating and that for each* $T \geq 0$, *there exists* $\mathbb{Q} \in \mathcal{M}_T^e$ *such that* $H(\mathbb{Q}; 0, T) < \infty$. *Then there exists* $\pi \in \mathcal{A}$ *such that*

$$(4.8) \qquad \frac{1}{\gamma_t} = \frac{1}{\gamma_0} + \int_0^t \pi_u \, dS_u \qquad \textit{for all } t \geq 0, \ a.s.$$

REMARK 4.8.

1. The additional condition that there exists an equivalent local martingale measure with finite entropy is standard in the literature. It corresponds to the existence of the primal optimizer in related exponential-utility maximization problems (see [8, 14, 18]). Such a condition would follow immediately if we assumed that the essential suprema in the definition of self-generating random fields were attained in the appropriate domain (see [1]). A simple sufficient condition for the existence of an equivalent finite-entropy local martingale measure will be given in Lemma 5.4 in Section 5 below.
2. Identification of sufficient conditions for condition 2(b) reduces to the stipulation that (4.8) holds for some $\pi \in \mathcal{A}_{\mathrm{bd}}$, together with the verification of the $\mathbb{Q}$-martingale property of the *local* martingale $(1/\gamma_t)_{t \in [0,T]}$, for all $T \geq 0$ and all $\mathbb{Q} \in \mathcal{M}_T^H$. A simple (and far from necessary) criterion is that $1/\gamma \in \mathbb{L}^\infty$, for all $t \geq 0$.

4.3. *On condition* 2(c) *of Theorem* 4.4. We turn to condition 2(c) of Theorem 4.4, assuming throughout that conditions 2(a) and 2(b) hold. For $T \geq 0$ and $\mathbb{Q} \in \mathcal{M}_T^H$, we define a measure $\mathbb{Q}_\gamma$ on $\mathcal{F}_T$ by

$$(4.9) \qquad \frac{d\mathbb{Q}_\gamma}{d(\mathbb{P}|_{\mathcal{F}_T})} = \frac{\gamma_0}{\gamma_T} \frac{d\mathbb{Q}}{d(\mathbb{P}|_{\mathcal{F}_T})}.$$

Thanks to condition 2(b), $\mathbb{Q}_\gamma$ is a well-defined probability measure. It is known in mathematical finance as the *forward measure* with respect to a numéraire-change $(1)_{t \in [0,T]} \to (\gamma_t)_{t \in [0,T]}$. It is notationally convenient to introduce the following set

$$\mathcal{M}_T^\gamma = \{\mathbb{Q}_\gamma : \mathbb{Q} \in \mathcal{M}_T^H\}.$$

Since $\frac{1}{\gamma_T} Z_T^{\mathbb{Q}}/Z_t^{\mathbb{Q}} = \frac{1}{\gamma_t} Z_T^{\mathbb{Q}_\gamma}/Z_t^{\mathbb{Q}_\gamma}$, the relative conditional entropy $H$ takes a particularly simple form when written in terms of $\mathbb{Q}_\gamma$:

$$H(\mathbb{Q}; t, T) = h\left(\frac{1}{\gamma_t}\right) - \frac{1}{\gamma_t}(\log(Z_t^{\mathbb{Q}_\gamma}) + \mathbb{E}^{\mathbb{Q}_\gamma}[\log(Z_T^{\mathbb{Q}_\gamma}) - A_T | \mathcal{F}_t]).$$

For $\mathbb{Q}_\gamma \in \mathcal{M}_T^\gamma$, define the process $(F_t^{\mathbb{Q}_\gamma})_{t\in[0,T]}$ as

$$F_t^{\mathbb{Q}_\gamma} = A_t - \log(Z_t^{\mathbb{Q}_\gamma}), \qquad t \in [0, T].$$

The following proposition is a simple consequence of Theorem 3.14.

PROPOSITION 4.9. *Suppose that conditions* 2(a) *and* 2(b) *of Theorem 4.4 hold. Then condition* 2(c) *is equivalent to the conjunction of the following two statements:*

1. *for each $T \geq 0$ and $\mathbb{Q}_\gamma \in \mathcal{M}_T^\gamma$, the process $(F_t^{\mathbb{Q}_\gamma})_{t\in[0,T]}$ is a $\mathbb{Q}_\gamma$-supermartingale and*
2. *for each $T \geq 0$ there exists $\hat{\mathbb{Q}}_\gamma \in \mathcal{M}_T^\gamma$ such that $(F_t^{\hat{\mathbb{Q}}_\gamma})_{t\in[0,T]}$ is a $\hat{\mathbb{Q}}_\gamma$-martingale.*

## 5. Itô-process models.

5.1. *The main result.* Having characterized exponential self-generating utility random fields in general locally-bounded semimartingale market models of Section 4, we turn to a specific class of models where we can say a great deal more.

Consider a special case of the financial model of Section 2 with one risky asset driven by a single Brownian motion $(B_t)_{t\in[0,\infty)}$ on a filtration generated by two independent Brownian motions $(B_t)_{t\in[0,\infty)}$ and $(W_t)_{t\in[0,\infty)}$. The price-process $(S_t)_{t\in[0,\infty)}$, defined on the underlying filtration $\mathbb{F} = (\mathcal{F}_t)_{t\in[0,\infty)}$—a natural augmentation of the filtration generated by $B$ and $W$—admits the following differential representation:

(5.1) $$dS_t = \theta_t \, dt + dB_t, \qquad t \geq 0, S_0 = s_0 \in \mathbb{R},$$

where $(\theta_t)_{t\in[0,\infty)}$ is an $\mathbb{F}$-progressively-measurable processes.

REMARK 5.1. Our choice of unit volatility and "arithmetic" evolution of the stock price entails no loss of generality compared to the models usually found in the literature; one can replicate exactly the same contingent claims. On the other hand, such a simplification relieves the notation and renders the central idea more transparent. Similarly, an extension to a model with several driving Brownian motions—and several assets—is straightforward and its treatment would only inflate the already heavy notation.

We assume that $\int_0^T \theta_u^2 \, du < \infty$, for all $T > 0$, a.s. and that the stochastic process $(Z_t^{\theta,0})_{t\in[0,\infty)}$, defined by

(5.2) $$Z_t^{\theta,0} = \exp\left(-\int_0^t \theta_u \, dB_u - \frac{1}{2}\int_0^t \theta_u^2 \, du\right), \qquad t \geq 0,$$

is a martingale on $[0, \infty)$, so that the condition NFLVRFH of Assumption 2.1 of Section 2.3 is satisfied.

The main result of this section is the following.

THEOREM 5.2. *Assume that the price process $(S_t)_{t \in [0,\infty)}$ is given by (5.1), let $(\gamma_t)_{t \in [0,\infty)}$ and $(A_t)_{t \in [0,\infty)}$ be stochastic processes and define the mapping $U : \Omega \times [0, \infty) \times \mathbb{R} \to \mathbb{R}$ as*

$$U(t, x) = -e^{-\gamma_t x + A_t}. \tag{5.3}$$

*If $U$ is a self-generating utility random field, and*

$$\forall p > 1 \quad \frac{1}{\gamma_T} \in \mathbb{L}^p(\mathcal{F}_T), \tag{5.4}$$

$$\exists \varepsilon > 0 \quad Z_T^{\theta,0} \in \mathbb{L}^{1+\varepsilon}(\mathcal{F}_T) \tag{5.5}$$

*hold for all $T \geq 0$, then:*

1. *both $\gamma$ and $A$ are continuous semimartingales, and*
2. *there exist progressively-measurable processes $(\delta_t)_{t \in [0,\infty)}$, $(\phi_t)_{t \in [0,\infty)}$ and $(\rho_t)_{t \in [0,\infty)}$ with $\int_0^T (\delta_u^2 + \phi_u^2 + \rho_u^2) \, du < \infty$, for all $T > 0$, a.s., such that for all $t \geq 0$, we have*

$$\frac{1}{\gamma_t} = \frac{1}{\gamma_0} + \int_0^t \frac{1}{\gamma_u} \delta_u \, dS_u, \tag{5.6}$$

*and*

$$A_t = A_0 + \frac{1}{2} \int_0^t (\theta_u - \delta_u)^2 \, du + \gamma_t \int_0^t \rho_u \, dS_u \tag{5.7}$$
$$- \frac{1}{2} \int_0^t \phi_u^2 \, du - \int_0^t \phi_u \, dW_u.$$

*Conversely, suppose that the processes $\gamma$ and $A$ are continuous semimartingales admitting representations (5.6) and (5.7), and, additionally, that the following regularity conditions are met for all $T \geq 0$:*

$$\forall n \in \mathbb{N} \quad \exp(A_T + n\gamma_T) \in \mathbb{L}^1(\mathcal{F}_T) \tag{5.8}$$

*and*

$$\sup_{t \in [0,T]} (|\delta_t| + |\rho_t| + |\phi_t|) \in \mathbb{L}^\infty(\mathcal{F}_T), \quad \mathbb{E}[e^{1/2 \int_0^T \theta_t^2 \, dt}] < \infty. \tag{5.9}$$

*Then $U$ is self-generating.*

REMARK 5.3. Thanks to Novikov's criterion and the Hölder's inequality that martingale property of $Z^{\theta,0}$ and assumption (5.5) are implied, for instance, by the following Novikov-type condition:

$$\forall T \geq 0, \exists \varepsilon > 0 \quad \mathbb{E}[e^{(1/2+\varepsilon) \int_0^T \theta_u^2 \, du}].$$

5.2. *Proof of Theorem 5.2.* Before we focus on the proof itself, we establish several auxiliary results. We choose a time-horizon $T > 0$ and keep it fixed throughout the proof. If a different time-horizon is needed, the reader will be explicitly warned.

5.2.1. *Martingale measures.* Let $\mathcal{P}$ denote the set of all $\mathbb{F}$-progressively measurable processes $(\nu_t)_{t \in [0,T]}$ such that $\int_0^T \nu_u^2 \, du < \infty$, a.s., and let

$$\mathcal{N} = \{(\nu_1, \nu_2) \in \mathcal{P} \times \mathcal{P} : Z^{\nu_1, \nu_2} \text{ is a true martingale}\},$$

where the positive local martingale $(Z_t^{\nu_1, \nu_2})_{t \in [0,T]}$ is given by

$$dZ_t^{\nu_1,\nu_2} = -Z_t^{\nu_1,\nu_2}(\nu_1(u)\, dB_u + \nu_2(u)\, dW_u), \qquad Z_0^{\nu_1,\nu_2} = 1.$$

For $(\nu_1, \nu_2) \in \mathcal{N}$, we define the probability measure $\mathbb{Q}^{\nu_1,\nu_2} \sim \mathbb{P}|_{\mathcal{F}_T}$ by

$$\frac{d\mathbb{Q}^{\nu_1,\nu_2}}{d(\mathbb{P}|_{\mathcal{F}_T})} = Z_T^{\nu_1,\nu_2}.$$

By virtue of Girsanov's theorem (see [11] for details), a probability measure $\mathbb{Q} \sim \mathbb{P}|_{\mathcal{F}_T}$ belongs to $\mathcal{M}_T^e$ if and only if there exists $\nu \in \mathcal{P}$ such that $(\theta, \nu) \in \mathcal{N}$ and $\frac{d\mathbb{Q}}{d(\mathbb{P}|_{\mathcal{F}_T})} = Z_T^{\theta,\nu}$. Let us introduce the following families:

$$\mathcal{P}^\infty = \Big\{\nu \in \mathcal{P} : \sup_{t \in [0,T]} |\nu_t| \in \mathbb{L}^\infty\Big\},$$
$$\mathcal{P}^{\nu_1} = \{\nu \in \mathcal{P} : (\nu_1, \nu) \in \mathcal{N}\} \qquad \text{for } \nu_1 \in \mathcal{P}.$$

LEMMA 5.4. *Suppose that the condition (5.5) holds and that the random field $U$ of (5.3) is self-generating. Then for each $\nu \in \mathcal{P}^\infty$,*

1. $(\theta, \nu) \in \mathcal{N}$, *and*
2. $H(\mathbb{Q}^{\theta,\nu}; 0, T) < \infty$.

*In particular, $\mathcal{M}_T^e \cap \mathcal{M}_T^H \neq \varnothing$.*

PROOF. We first show that $\mathbb{Q}^{\theta,0} \in \mathcal{M}_T^H$. Hölder's inequality used in conjunction with assumptions (5.4) and (5.5) yields $(\frac{1}{\gamma_T} Z_T^{\theta,0})^{1+\varepsilon} \in \mathbb{L}^1(\mathcal{F}_T)$, for small enough $\varepsilon > 0$. Moreover, since $Z_T^{\theta,\nu}/Z_T^{\theta,0} \in \bigcap_{p>1} \mathbb{L}^p(\mathcal{F}_T)$ for $\nu \in \mathcal{P}^\infty$, we have $(\frac{1}{\gamma_T} Z_T^{\theta,\nu})^{1+\varepsilon} \in \mathbb{L}^1(\mathcal{F}_T)$, for small enough $\varepsilon > 0$. Consequently,

$$(5.10) \qquad h\left(\frac{1}{\gamma_T} Z_T^{\theta,\nu}\right) \in \mathbb{L}^1(\mathcal{F}_T) \qquad \forall \nu \in \mathcal{P}^\infty.$$

Using the elementary inequality $xy \leq x \log x - x + e^y$ for all $x \geq 0$ and $y \in \mathbb{R}$, condition (5.5) implies

$$(5.11) \qquad \frac{1}{\gamma_T} Z_T^{\theta,\nu} A_T \in \mathbb{L}^1(\mathcal{F}_T).$$

Assertions (5.10) and (5.11) yield $H(\mathbb{Q}^{\theta,\nu}; 0, T) < \infty$, for all $\nu \in \mathcal{P}^\infty$. □

5.2.2. *Proof of necessity.* We first show that the self-generation property of $U$ implies (5.6). By Theorem 4.4, the processes $\frac{1}{\gamma}$ and $A$ are semimartingales and by Lemma 5.4 and Proposition 4.7, for each $T \geq 0$, there exists a progressive process $(\hat{\delta}_t^{(T)})_{t \in [0,T]}$ such that

$$\text{(5.12)} \qquad \frac{1}{\gamma_T} = \frac{1}{\gamma_0} + \int_0^T \hat{\delta}_u^{(T)} \, dS_u.$$

For $0 \leq T_1 < T_2$, processes $\hat{\delta}^{(T_1)}$ and $\hat{\delta}^{(T_2)}$ agree $d\mathbb{P} \times dt$ on $\Omega \times [0, T_1]$. So, there exists a progressively measurable process $(\hat{\lambda}_t)_{t \in [0,\infty)}$ such that

$$\text{(5.13)} \qquad \frac{1}{\gamma_\cdot} = \frac{1}{\gamma_0} + \int_0^\cdot \hat{\lambda}_u \, dS_u \qquad \text{a.s.}$$

Finally, positivity of $\gamma$ implies (5.6) with $\delta_t = \gamma_t \hat{\delta}_t$.

Next, we turn to the process $A$. To better understand its structure, we construct a fictional constrained financial market, as a technical tool. It comprises of three securities $\tilde{S} = (\tilde{S}^0, \tilde{S}^1, \tilde{S}^2)$, given by

$$\text{(5.14)} \qquad \begin{cases} \tilde{S}_t^0 = B_t + \int_0^t \tilde{\theta}_u \, du & \text{where } \tilde{\theta}_t = \theta_t - \delta_t, \\ \tilde{S}_t^1 = t \quad \text{and} \\ \tilde{S}_t^2 = W_t, \end{cases}$$

with portfolios $\pi = (\pi^0, \pi^1, \pi^2)$, representing the numbers of shares of each of the three securities, constrained to take values in the convex set

$$\text{(5.15)} \qquad K = \{(\pi^0, \pi^1, \pi^2) \subseteq \mathbb{R}^3 : \pi^1 + \tfrac{1}{2}(\pi^2)^2 \leq 0\}.$$

Let $\mathcal{A}^K$ denote the set of 3-dimensional $\tilde{S}$-integrable processes $(\pi_t)_{t \in [0,\infty)}$ with $\pi_t \in K$ for all $t \in [0,T]$, a.s.

The central argument in the proof below is based on a version of the Optional Decomposition theorem. For the reader's convenience, we rephrase the pertinent content of Theorem 3.1 in [12] in our setting, noting that its technical conditions are satisfied thanks to Proposition B.1 which establishes closedness with respect to the semimartingale topology of the family

$$\tilde{\mathcal{S}} := \left\{ \int_0^\cdot \pi_u \, d\tilde{S}_u : \pi \in \mathcal{A}^K \right\}.$$

THEOREM 5.5 (Föllmer and Kramkov, 1997). *Let $(V_t)_{t \in [0,T]}$ be a càdlàg and adapted process which is locally bounded from below. Then the following statements are equivalent:*

1. *$V$ has a decomposition of the form*

$$V_t = V_0 + (\pi \cdot \tilde{S})_t + D_t, \qquad t \in [0, T],$$

*for some portfolio process $\pi \in \mathcal{A}^K$ and a nonincreasing adapted càdlàg process $(D_t)_{t \in [0,T]}$, with $D_0 = 0$.*

2. $V_t - A_t^{\mathbb{Q}}$ is a $\mathbb{Q}$-local supermartingale for all $\mathbb{Q} \in \tilde{\mathcal{M}}$, where $\tilde{\mathcal{M}}$ is the set of all probability measures $\mathbb{Q}$ on $\mathcal{F}_T$ with $\mathbb{Q} \sim \mathbb{P}|_{\mathcal{F}_T}$ such that the $\mathbb{Q}$-compensators of the wealth processes $\pi \cdot \tilde{S}$ are bounded, in the sense of positive measures on the optional sets, uniformly over all admissible $\pi \in \mathcal{A}^K$ (the process $A_t^{\mathbb{Q}}$ denotes their least upper bound).

Let us first identify the set $\tilde{\mathcal{M}}$ and the processes $A^{\mathbb{Q}}$ for $\mathbb{Q} \in \tilde{\mathcal{M}}$ in our market $\tilde{S}$. A generic wealth process $\pi \cdot \tilde{S}$ has the following differential representation under a generic measure $\mathbb{Q}^{\nu_1,\nu_2}$:

$$d(\pi \cdot \tilde{S})_t = (\pi^0(t)\tilde{\theta}_t - \pi^0(t)\nu_1(t) + \pi^1(t) + \pi^2(t)\nu_2(t))\,dt + dL_t$$

for some $\mathbb{Q}^{\nu_1,\nu_2}$-local martingale $L$. Note that for a fixed $t \in [0,T]$,

$$\sup_{(\pi^0,\pi^1,\pi^2) \in K} (\pi^0\tilde{\theta}_t - \pi^0\nu_1(t) + \pi^1 + \pi^2\nu_2(t)) = \begin{cases} \infty, & \nu_1(t) \neq \tilde{\theta}_t, \\ -\tfrac{1}{2}\nu_2^2(t), & \nu_1(t) = \tilde{\theta}_t. \end{cases}$$

Therefore,

$$\tilde{\mathcal{M}} = \{\mathbb{Q}^{\tilde{\theta},\nu} : \nu \in \mathcal{P}^{\tilde{\theta}}\} \quad \text{and} \quad A_t^{\mathbb{Q}^{\tilde{\theta},\nu}} = \frac{1}{2}\int_0^t \nu_u^2\,du.$$

Thanks to the choice of the coefficient $\tilde{\theta}$ and the already proven relation (5.6), we have $\frac{d\mathbb{Q}^{\tilde{\theta},\nu}}{d\mathbb{P}} = \frac{\gamma_0}{\gamma_T}\frac{d\mathbb{Q}^{\theta,\nu}}{d\mathbb{P}}$, whenever $\mathbb{E}^{\mathbb{Q}^{\theta,\nu}}[\frac{\gamma_0}{\gamma_T}] = 1$. By condition 2(b) of Theorem 4.4, this equality holds for all $\nu \in \mathcal{P}^{\theta}$, such that $\mathbb{Q}^{\theta,\nu} \in \mathcal{M}_T^H$. In particular, by Lemma 5.4, it holds for $\nu \in \mathcal{P}^{\infty}$. In the notation of Section 4.3, we have $\mathbb{Q}^{\tilde{\theta},\nu} = (\mathbb{Q}^{\theta,\nu})_\gamma$, i.e., $\mathbb{Q}^{\tilde{\theta},\nu}$ is the forward measure associated with $\mathbb{Q}^{\theta,\nu}$ and the numéraire $\gamma$.

By Proposition 4.9, the process $F^{\mathbb{Q}^{\tilde{\theta},\nu}} = A - \log(Z^{\tilde{\theta},\nu})$ is a $\mathbb{Q}^{\tilde{\theta},\nu}$-supermartingale for each $\nu$ such that $\mathbb{Q}^{\tilde{\theta},\nu} \in \mathcal{M}_T^H$. So, by Lemma 5.4, it is a $\mathbb{Q}^{\tilde{\theta},\nu}$-supermartingale for all $\nu \in \mathcal{P}^{\infty}$. Thus, for an arbitrary $\nu \in \mathcal{P}^{\tilde{\theta}}$, the process $F^{\mathbb{Q}^{\tilde{\theta},\nu^n}}$ is a $\mathbb{Q}^{\tilde{\theta},\nu^n}$-supermartingale, for each $n \in \mathbb{N}$, where

$$\nu_t^n = \nu_t \mathbf{1}_{\{t \leq \tau_n\}} \qquad \text{where } \tau_n = \inf\left\{t \geq 0 \colon \int_0^t \nu_u^2\,du > n\right\}.$$

Processes $F^{\mathbb{Q}^{\tilde{\theta},\nu^n}}$ and $F^{\mathbb{Q}^{\tilde{\theta},\nu}}$, as well as measures $\mathbb{Q}^{\tilde{\theta},\nu^n}$ and $\mathbb{Q}^{\tilde{\theta},\nu}$, agree on $[0,\tau_n]$. So, the stopped process $(F^{\mathbb{Q}^{\tilde{\theta},\nu}})^{\tau_n}$ is a $\mathbb{Q}^{\tilde{\theta},\nu}$-supermartingale, for all $n \in \mathbb{N}$ and $\mathbb{P}[\tau_n \leq T] \to 0$. In other words, $(F_t^{\mathbb{Q}})_{t \in [0,T]}$ is a local $\mathbb{Q}$-supermartingale, for each $\mathbb{Q} \in \tilde{\mathcal{M}}$. A simple application of Itô's formula implies that the process

$$A_t - \frac{1}{2}\int_0^t \tilde{\theta}_u^2\,du - \frac{1}{2}\int_0^t \nu_u^2\,du, \qquad t \in [0,T],$$

is also a local $\mathbb{Q}^{\tilde{\theta},\nu}$-supermartingale, for each $\nu$ with $\mathbb{Q}^{\tilde{\theta},\nu} \in \mathcal{M}_T^H$. Theorem 5.5 yields the existence of a portfolio process $\pi \in \mathcal{A}^K$ and a nonincreasing adapted càdlàg process $(D_t)_{t\in[0,T]}$ such that

$$A_t = A_0 + \int_0^t \tilde{\pi}_u \, d\tilde{S}_u + D_t.$$

Thanks to part 2 of Proposition 4.9, the process $D$ must vanish identically. For the same reason, there can be no "slack" in the portfolio process $\tilde{\pi}$, i.e., $\pi^1(t) = -\frac{1}{2}\pi^2(t)^2$, $d\mathbb{P} \times dt$-a.e. Consequently, with $\phi = -\pi^2$, the process $A$ has the following form:

$$A_t = A_0 + \frac{1}{2}\int_0^t \tilde{\theta}_u^2 \, du + \int_0^t \pi_u^0 (dB_u + \tilde{\theta}_u) - \frac{1}{2}\int_0^t \phi_u^2 \, du - \int_0^t \phi_u \, dW_u.$$

Let the process $\rho$ be defined as $\rho_t = \delta_t \hat{\rho}_t + \frac{1}{\gamma_t}\pi_t^0$, where $\hat{\rho}_t = \frac{1}{\gamma_t}\int_0^t \pi^0 \, d\tilde{S}_u$. A straightforward calculation using the identity $d(\gamma_t \int_0^t \zeta_t \, dS_t) = \gamma_t(\zeta_t - \delta_t \times \int_0^t \zeta_u \, dS_u)(\tilde{\theta}_t \, dt + dB_t)$ which holds for any $\zeta \in \mathcal{P}$ implies (5.7), for a fixed $T$. Finally, the argument for the passage from (5.12) to (5.13) can be reused to show the validity of (5.7) on the whole positive semi-axis.

5.2.3. *Proof of sufficiency.* The proof of sufficiency is based on Theorem 4.4. Condition 2(a) of Theorem 4.4 is assumed in (5.8). Boundedness of the process $\delta$ ensures that $\frac{1}{\gamma}$ is a $\mathbb{Q}^{\theta,\nu}$-martingale for all $\nu \in \mathcal{P}^\theta$. In particular, condition 2(b) of Theorem 4.4 is fulfilled. To verify condition 2(c) of Theorem 4.4, we turn to the characterization in Proposition 4.9: the process $F^{\mathbb{Q}^{\tilde{\theta},\nu}}$ of Section 4.3 can be written as

$$F_t^{\mathbb{Q}^{\tilde{\theta},\nu}} = M_t - \int_0^t \phi_u \, d\tilde{W}_u - \frac{1}{2}\int_0^t (\nu_u - \phi_u)^2 \, du,$$

where $M_t = \gamma_t \int_0^t \rho \, dS_u$ and $\tilde{W}_t = W_t + \int_0^t \nu_u \, du$. Thanks to (5.9), processes $M_t$ and $\int_0^t \phi_u \, d\tilde{W}_u$ are martingales under the forward measure $\mathbb{Q}^{\tilde{\theta},\nu}$. So, statement 1 of Proposition 4.9 holds. In order to verify statement 2, we take $\nu = \phi$, noting that $Z^{\theta,\phi}$ is a true martingale. By Proposition 4.9, $U$ is a self-generating utility random field.

## APPENDIX A: CONVEX DUALITY FOR RANDOM FIELDS

For the purposes of this section, we fix $0 \leq t \leq T < \infty$ and a random variable $\kappa \in \mathbb{L}_+^\infty(\mathcal{F}_t)$. Unless designated otherwise, all the $\mathbb{L}^p$-spaces (and their duals), $p \in [0,\infty]$, will be with respect to $(\Omega, \mathcal{F}_T, \mathbb{P}|_{\mathcal{F}_T})$. The space $\mathbb{L}^1$ will always be identified with its image in $(\mathbb{L}^\infty)^*$ under the canonical isometric embedding of a Banach space into its bidual.

We (re-)introduce the following variations of the standard notation:

1. The functional $\mathbb{U}_\kappa(\cdot):\mathbb{L}^\infty \to \mathbb{R}$ is defined by $\mathbb{U}_\kappa(\zeta) = \mathbb{E}[\kappa U(T,\zeta)]$, for $\zeta \in \mathbb{L}^\infty$.
2. The convex conjugate $\mathbb{V}_\kappa:(\mathbb{L}^\infty)^* \to (-\infty,\infty]$ of $\mathbb{U}_\kappa$ is given by
$$\mathbb{V}_\kappa(\zeta^*) = \sup_{\zeta \in \mathbb{L}^\infty}(\mathbb{U}_\kappa(\zeta) - \langle \zeta^*, \zeta \rangle) \qquad \text{for } \zeta^* \in (\mathbb{L}^\infty)^*.$$
3. $\mathcal{K}_{t\to T} = \{\int_t^T \pi_u\, dS_u : \pi \in \mathcal{A}_{\mathrm{bd}}\}$.
4. $\mathcal{C}_{t\to T} = (\mathcal{K}_{t\to T} - \mathbb{L}^0_+) \cap \mathbb{L}^\infty$.
5. $\mathcal{D}_{t\to T} = \{\zeta^* \in (\mathbb{L}^\infty)^* : \langle \zeta^*, \zeta \rangle \le 0 \text{ for all } \zeta \in \mathcal{C}_{t\to T}\}$.
6. $u(\xi; t, T) = \operatorname*{ess\,sup}_{\pi \in \mathcal{A}_{\mathrm{bd}}} \mathbb{E}[U(T, \xi + \int_t^T \pi_u\, dS_u)|\mathcal{F}_t]$, for $\xi \in \mathbb{L}^\infty(\mathcal{F}_t)$.
7. $v(\eta; t, T) = \operatorname*{ess\,inf}_{\mathbb{Q} \in \mathcal{M}_T^a} \mathbb{E}[V(T, \eta Z_T^\mathbb{Q}/Z_t^\mathbb{Q})|\mathcal{F}_t]$, for $\eta \in \mathbb{L}^1_+(\mathcal{F}_t)$.
8. $u_\kappa(\zeta) = \sup_{\rho \in \mathcal{C}_{t\to T}} \mathbb{U}_\kappa(\zeta + \rho) \in (-\infty,\infty]$, for $\zeta \in \mathbb{L}^\infty$.
9. $\mathcal{D}_{t\to T}^\eta = \{\zeta^* \in \mathcal{D}_{t\to T} : \langle \zeta^*, \xi \rangle = \langle \eta, \xi \rangle \text{ for all } \xi \in \mathbb{L}^\infty(\mathcal{F}_t)\}$, for $\eta \in \mathbb{L}^1_+(\mathcal{F}_t)$.
10. $v_\kappa(\eta) = \inf_{\zeta^* \in \mathcal{D}_{t\to T}^\eta} \mathbb{V}_\kappa(\zeta^*)$, for $\eta \in \mathbb{L}^1_+(\mathcal{F}_t)$ and $v_\kappa(\eta) = \infty$, for $\eta \in \mathbb{L}^1(\mathcal{F}_t) \setminus \mathbb{L}^1_+(\mathcal{F}_t)$.

PROPOSITION A.1. *For $\zeta_0 \in \mathbb{L}^\infty$, we have*
$$(A.1) \qquad u_\kappa(\zeta_0) = \inf_{\zeta^* \in \mathcal{D}_{t\to T}} (\mathbb{V}_\kappa(\zeta^*) + \langle \zeta^*, \zeta_0 \rangle),$$
*where the infimum above is attained at some $\hat{\zeta}^* \in \mathcal{D}_{t\to T}$.*

PROOF. Suppose first that $u_\kappa(\zeta_0) = \infty$. The definitions of $\mathbb{V}_\kappa$ and $\mathcal{D}_{t\to T}$ above ensure that for $\zeta^* \in \mathcal{D}_{t\to T}$ and $\rho \in \mathcal{C}_{t\to T}$, we have
$$\mathbb{V}_\kappa(\zeta^*) \ge \mathbb{U}_\kappa(\zeta_0 + \rho) - \langle \zeta^*, \zeta_0 + \rho \rangle \ge \mathbb{U}_\kappa(\zeta_0 + \rho) - \langle \zeta^*, \zeta_0 \rangle.$$
Taking a supremum of the right-hand side over all $\rho \in \mathcal{C}_{t\to T}$ implies that $\mathbb{V}_\kappa(\zeta^*) = \infty$ for all $\zeta^* \in \mathcal{D}_{t\to T}$, which, in turn, implies (A.1).

When $u_\kappa(\zeta_0) < \infty$, we define the following two subsets of $\mathbb{L}^\infty \times \mathbb{R}$:
$$A = \{(\zeta, u) \in \mathbb{L}^\infty \times \mathbb{R} : u \le u_\kappa(\zeta_0 + \zeta)\},$$
$$B = \mathcal{C}_{t\to T} \times [u_\kappa(\zeta_0), \infty).$$
It is straightforward to check that:

1. both $A$ and $B$ are convex and nonempty,
2. $\operatorname{Int} B \ne \varnothing$ (since $-\mathbb{L}^\infty_+ \subset \mathcal{C}_{t\to T}$), and
3. $A \cap \operatorname{Int} B = \varnothing$.

By the Hahn–Banach theorem (see Theorem 5.50, page 190 in [2]) there exists a constant $c \in \mathbb{R}$ and a nonnull element $(\hat{\zeta}^*, \hat{u}^*)$ of the dual space $(\mathbb{L}^\infty)^* \times \mathbb{R} \cong (\mathbb{L}^\infty \times \mathbb{R})^*$ such that
$$(A.2) \qquad \langle \hat{\zeta}^*, \zeta \rangle + u\hat{u}^* + c \ge 0$$
$$\forall (\zeta, u) \text{ such that } u \le \mathbb{U}_\kappa(\zeta_0 + \zeta + \rho), \text{ for some } \rho \in \mathcal{C}_{t\to T}$$

and

(A.3) $\qquad \langle \hat{\zeta}^*, \rho \rangle + u\hat{u}^* + c \leq 0 \qquad \forall u \geq u_\kappa(\zeta_0), \forall \rho \in \mathcal{C}_{t \to T}.$

From (A.3) and the fact that $0 \in \mathcal{C}_{t \to T}$, we conclude that $\hat{u}^* \leq 0$. Using (A.3) again, this time in conjunction with the positive homogeneity of $\mathcal{C}_{t \to T}$, we get $\langle \hat{\zeta}^*, \rho \rangle \leq 0$, for all $\rho \in \mathcal{C}_{t \to T}$, which, in turn, implies that $\hat{\zeta}^* \in \mathcal{D}_{t \to T} \subseteq (\mathbb{L}^\infty)^*_+$.

Our next task is to show that $\hat{u}^* < 0$. Suppose, to the contrary, that $\hat{u}^* = 0$. Then (A.2) and (A.3) imply that $\langle \hat{\zeta}^*, \zeta \rangle = c$ for all $\zeta$ in the intersection $\pi_{\mathbb{L}^\infty}(A) \cap \pi_{\mathbb{L}^\infty}(B)$ of the projections of $A$ and $B$ onto $\mathbb{L}^\infty$. Finiteness of $\mathbb{U}_\kappa(\zeta)$ for all $\zeta \in \mathbb{L}^\infty$ yields $\pi_{\mathbb{L}^\infty}(A) \cap \pi_{\mathbb{L}^\infty}(B) = \mathcal{C}_{t \to T}$. Thus, $\langle \hat{\zeta}^*, \zeta \rangle = c$, for all $\zeta \in \mathcal{C}_{t \to T}$. Since $-\mathbb{L}^\infty_+ \subseteq \mathcal{C}_{t \to T}$, this can only happen if $\hat{\zeta}^* = 0$, which is in contradiction with the assumptions that $\hat{u}^* = 0$ and the nontriviality of the separating functional $(\hat{\zeta}^*, \hat{u}^*)$.

Having established that $\hat{u}^* < 0$, we can assume, without loss of generality, that $\hat{u}^* = -1$. The equation (A.3) with $\rho = 0$ and $u = u_\kappa(\zeta_0)$ implies that $u_\kappa(\zeta_0) \geq c$. On the other hand, (A.2) states that

(A.4) $\qquad c \geq \mathbb{U}_\kappa(\zeta_0 + \zeta + \rho) - \langle \hat{\zeta}^*, \zeta \rangle \qquad \forall \zeta \in \mathbb{L}^\infty, \forall \rho \in \mathcal{C}_{t \to T}.$

The fact that $\langle \hat{\zeta}^*, \rho \rangle \leq 0$, for all $\rho \in \mathcal{C}_{t \to T}$, allows us to combine the previous conclusions with (A.4) to get the inequality

$$u_\kappa(\zeta_0) \geq \mathbb{U}_\kappa(\zeta) - \langle \hat{\zeta}^*, \zeta \rangle + \langle \hat{\zeta}^*, \zeta_0 \rangle + \langle \hat{\zeta}^*, \rho \rangle$$
$$\geq \mathbb{U}_\kappa(\zeta) - \langle \hat{\zeta}^*, \zeta \rangle + \langle \hat{\zeta}^*, \zeta_0 \rangle.$$

Taking the supremum over all $\zeta \in \mathbb{L}^\infty$, we obtain

(A.5) $\begin{aligned} u_\kappa(\zeta_0) &\geq \sup_{\zeta \in \mathbb{L}^\infty} (\mathbb{U}_\kappa(\zeta) - \langle \hat{\zeta}^*, \zeta \rangle) + \langle \hat{\zeta}^*, \zeta_0 \rangle = \mathbb{V}_\kappa(\hat{\zeta}^*) + \langle \hat{\zeta}^*, \zeta_0 \rangle \\ &\geq \inf_{\zeta^* \in \mathcal{D}_{t \to T}} (\mathbb{V}_\kappa(\zeta^*) + \langle \zeta^*, \zeta_0 \rangle). \end{aligned}$

On the other hand, by the definition of $\mathbb{V}_\kappa$, we have

$$\mathbb{U}_\kappa(\zeta_0 + \rho) \leq \mathbb{V}_\kappa(\zeta^*) + \langle \zeta^*, \zeta_0 + \rho \rangle \leq \mathbb{V}_\kappa(\zeta^*) + \langle \zeta^*, \zeta_0 \rangle$$

for all $\zeta^* \in \mathcal{D}_{t \to T}$ and all $\rho \in \mathcal{C}_{t \to T}$. Maximization of the left-hand side over all $\rho \in \mathcal{C}_{t \to T}$ and minimization of the right-hand side over all $\zeta^* \in \mathcal{D}_{t \to T}$ yield

(A.6) $\qquad u_\kappa(\zeta_0) \leq \inf_{\zeta^* \in \mathcal{D}_{t \to T}} (\mathbb{V}_\kappa(\zeta^*) + \langle \zeta^*, \zeta_0 \rangle).$

One only needs to combine (A.5) and (A.6) to finish the proof. $\square$

COROLLARY A.2. *For every $\eta \in \mathbb{L}^1_+(\mathcal{F}_t)$, we have*
$$v_\kappa(\eta) = \sup_{\xi \in \mathbb{L}^\infty(\mathcal{F}_t)} (u_\kappa(\xi) - \langle \eta, \xi \rangle).$$

PROOF. Proposition A.1 implies that $u_\kappa : \mathbb{L}^\infty(\mathcal{F}_t) \to (-\infty, \infty]$ is the convex conjugate of $v_\kappa : \mathbb{L}^1(\mathcal{F}_t) \to (-\infty, \infty]$, with respect to the pairing $(\xi, \eta) \mapsto \langle \xi, \eta \rangle = \mathbb{E}[\xi \eta]$ between $\mathbb{L}^\infty(\mathcal{F}_t)$ and $\mathbb{L}^1(\mathcal{F}_t)$. In order to complete the proof, we need to show that $v_\kappa$ is the convex conjugate of $u_\kappa$. It suffices to show that $v_\kappa$ is convex and lower semi-continuous with respect to the weak topology $\sigma(\mathbb{L}^1, \mathbb{L}^\infty)$ (see, e.g., Proposition 4.1, page 18 in [10]). For convexity, let $\varepsilon > 0$, $\alpha \in (0,1)$ and $\eta_1, \eta_2 \in \mathbb{L}^1_+$, and choose $\zeta_1^* \in \mathcal{D}_{t \to T}^{\eta_1}$ and $\zeta_2^* \in \mathcal{D}_{t \to T}^{\eta_2}$ such that $\mathbb{V}_\kappa(\zeta_1^*) \leq v_\kappa(\eta_1) + \varepsilon/2$ and $\mathbb{V}_\kappa(\zeta_2^*) \leq v_\kappa(\eta_2) + \varepsilon/2$. Then, by convexity of $\mathbb{V}_\kappa$, we have

$$\alpha v_\kappa(\eta_1) + (1-\alpha) v_\kappa(\eta_2) \geq -\varepsilon + \alpha \mathbb{V}_\kappa(\zeta_1^*) + (1-\alpha) \mathbb{V}_\kappa(\zeta_2^*)$$
$$\geq -\varepsilon + \mathbb{V}_\kappa(\alpha \zeta_1^* + (1-\alpha) \zeta_2^*).$$

It is straightforward to show that $\alpha \zeta_1^* + (1-\alpha) \zeta_2^* \in \mathcal{D}_{t \to T}^{\alpha \eta_1 + (1-\alpha) \eta_2}$ and conclude that $v_\kappa$ is, indeed, convex.

To establish lower semi-continuity, we take a directed set $A$ and a net $(\eta_\alpha)_{\alpha \in A}$ in $\mathbb{L}^1$ with $\eta_\alpha \to \eta$ weakly, and aim to show that $v_\kappa(\eta) \leq \liminf_\alpha v_\kappa(\eta_\alpha)$. Without loss of generality, we assume that $\eta_\alpha \in \mathbb{L}^1_+$ and $v_\kappa(\eta_\alpha) < \infty$, for all $\alpha \in A$, and that the limit $\lim_\alpha v_\kappa(\eta_\alpha)$ exists in $(-\infty, \infty]$. Let $(\varepsilon_\alpha)_{\alpha \in A}$ be a net in $(0, \infty)$ converging to 0, and let $(\zeta_\alpha^*)_{\alpha \in A}$ be a net in $\mathcal{D}_{t \to T}$ with $\zeta_\alpha^* \in \mathcal{D}_{t \to T}^{\eta_\alpha}$ such that $v_\kappa(\eta_\alpha) \geq \mathbb{V}_\kappa(\zeta_\alpha^*) - \varepsilon_\alpha$. By the Banach–Alaoglu theorem, there exist a subnet of $(\zeta_\alpha^*)_{\alpha \in A}$ (which we do not relabel) and $\zeta^* \in (\mathbb{L}^\infty)^*_+$ such that $\zeta_\alpha^* \to \zeta^*$. By the weak-* closedness of $\mathcal{D}_{t \to T}$, we have $\zeta^* \in \mathcal{D}_{t \to T}$. We claim that $\zeta^* \in \mathcal{D}_{t \to T}^\eta$. Indeed, for $\xi \in \mathbb{L}^\infty(\mathcal{F}_t)$, we have

$$\langle \zeta^*, \xi \rangle = \lim_\alpha \langle \zeta_\alpha^*, \xi \rangle = \lim_\alpha \langle \eta_\alpha, \xi \rangle = \langle \eta, \xi \rangle.$$

By the weak-* lower semi-continuity of $\mathbb{V}_\kappa$ (guaranteed by its definition as conjugate functional), we have

$$v_\kappa(\eta) \leq \mathbb{V}_\kappa(\zeta^*) \leq \liminf_\alpha \mathbb{V}_\kappa(\zeta_\alpha^*) \leq \liminf_\alpha (v_\kappa(\eta_\alpha) + \varepsilon_\alpha) = \lim_\alpha v_\kappa(\eta_\alpha). \quad \square$$

PROPOSITION A.3. *The following representation holds for any* $\zeta^* \in \mathcal{D}_{t \to T}$

$$\mathbb{V}_\kappa(\zeta^*) = \begin{cases} \mathbb{E}\left[\kappa V\left(T, \frac{1}{\kappa} \zeta^*\right)\right], & \zeta^* \in \mathbb{L}^1_+ \text{ and } \{\zeta^* > 0\} \subseteq \{\kappa > 0\}, \\ \infty, & \text{otherwise.} \end{cases}$$

PROOF. We divide the proof into several cases, depending on the "region" in which $\zeta^*$ lies:

1. $\zeta^*$ *is not in* $(\mathbb{L}^\infty)^*_+$. Then there exists $\zeta \in \mathbb{L}^\infty_+$ such that $M = \langle \zeta^*, \zeta \rangle < 0$. By monotonicity, $\mathbb{U}_\kappa(n\zeta) \geq \mathbb{U}_\kappa(0)$ for all $n \in \mathbb{N}$. So,

$$\mathbb{V}_\kappa(\zeta^*) \geq \limsup_{n \in \mathbb{N}} (\mathbb{U}_\kappa(n\zeta) - n \langle \zeta^*, \zeta \rangle)$$

$$\geq \limsup_{n \in \mathbb{N}} (\mathbb{U}_\kappa(0) + n |M|) = \infty.$$

2. $\zeta^*$ *is in* $(\mathbb{L}^\infty)^*_+$ *but not in* $\mathbb{L}^1_+$. The mapping $\mu_{\zeta^*}: \mathcal{F}_T \to [0,1]$, defined by $\mu_{\zeta^*}(A) = \langle \zeta^*, \mathbf{1}_A \rangle$, $A \in \mathcal{F}_T$, is a finitely-additive probability on $\mathcal{F}_T$. The condition that $\zeta^* \notin \mathbb{L}^1_+$ implies that $\mu_{\zeta^*}$ is not countably-additive. Thus, there exist a constant $\varepsilon > 0$ and a nonincreasing sequence $\{A_n\}_{n \in \mathbb{N}}$ of events in $\mathcal{F}_T$ such that $\bigcap_n A_n = \varnothing$ and $\langle \zeta^*, \mathbf{1}_{A_n} \rangle \geq \varepsilon$, for all $n \in \mathbb{N}$. Let $\{a_n\}_{n \in \mathbb{N}}$ be as in Definition 3.3 and let the sequence $\{\zeta_n\}_{n \in \mathbb{N}}$ in $\mathbb{L}^\infty$ be given by $\zeta_n = -a_n \mathbf{1}_{A_n}$. Due to nonsingularity of $U$,

$$\mathbb{V}_\kappa(\zeta^*) \geq \limsup_{n \in \mathbb{N}}(\mathbb{U}_\kappa(\zeta_n) - \langle \zeta^*, \zeta_n \rangle)$$

$$\geq \limsup_{n \in \mathbb{N}}(\mathbb{U}_\kappa(\zeta_n) + a_n \langle \zeta^*, \mathbf{1}_{A_n} \rangle)$$

$$\geq \limsup_{n \in \mathbb{N}} a_n \left( \frac{1}{a_n} \mathbb{U}_\kappa(-a_n \mathbf{1}_{A_n}) + \varepsilon \right) = \infty.$$

3. $\zeta^*$ *is in* $\mathbb{L}^1_+$ *and* $\mathbb{P}[\{\zeta^* > 0\} \cap \{\kappa = 0\}] > 0$. For $n \in \mathbb{N}$, define $\zeta_n = -n \mathbf{1}_A$, where $A = \{\zeta^* > 0\} \cap \{\kappa = 0\}$. Then $\mathbb{U}_\kappa(\zeta_n) = \mathbb{E}[\kappa U(T, \zeta_n)] = \mathbb{E}[\kappa U(T, 0))] = \mathbb{U}_\kappa(0)$, so

$$\mathbb{V}_\kappa(\zeta^*) \geq \limsup_{n \in \mathbb{N}}(\mathbb{U}_\kappa(\zeta_n) - \langle \zeta^*, \zeta_n \rangle)$$

$$\geq \limsup_{n \in \mathbb{N}}(\mathbb{U}_\kappa(0) + n \mathbb{E}[\zeta^* \mathbf{1}_A]) = \infty.$$

4. $\zeta^*$ *is in* $\mathbb{L}^1_+$ *and* $\{\zeta^* > 0\} \subseteq \{\kappa > 0\}$. For any $\zeta \in \mathbb{L}^\infty$, we have $\kappa \zeta \frac{1}{\kappa} \zeta^* = \zeta \zeta^*$, and so,

$$\kappa U(T, \zeta) \leq \kappa \zeta \frac{1}{\kappa} \zeta^* + \kappa V\left(T, \frac{1}{\kappa} \zeta^*\right) = \zeta \zeta^* + \kappa V\left(T, \frac{1}{\kappa} \zeta^*\right) \quad \text{a.s.}$$

for all $\zeta \in \mathbb{L}^\infty$. Therefore, $\mathbb{V}_\kappa(\zeta^*) \leq \mathbb{E}[\kappa V(T, \frac{1}{\kappa} \zeta^*)]$. To prove the opposite inequality, let $\{\zeta_n\}_{n \in \mathbb{N}}$ be given by

$$\zeta_n = -V'\left(T, \frac{1}{\kappa} \zeta^*\right) \mathbf{1}_{B_n},$$

where $B_n = \{\kappa > 0\} \cap \{-n \leq -V'(T, \frac{1}{\kappa} \zeta^*) \leq n\}$, so that $\zeta_n \in \mathbb{L}^\infty$. Then

$$\kappa U(T, \zeta_n) - \zeta_n \zeta^* = \kappa U(T, 0) \mathbf{1}_{B_n^c} + \kappa V\left(T, \frac{1}{\kappa} \zeta^*\right) \mathbf{1}_{B_n}.$$

The random variable $\kappa V(T, \frac{1}{\kappa} \zeta^*)$ is bounded from below by an integrable random variable [one can take $\kappa U(T, 0)$, for example]. So, the monotone convergence theorem implies that

$$\mathbb{E}[\kappa U(T, \zeta_n) - \zeta_n \zeta^*] \to \mathbb{E}\left[\kappa V\left(T, \frac{1}{\kappa} \zeta^*\right)\right],$$

which, in turn, yields $\mathbb{V}_\kappa(\zeta^*) \geq \mathbb{E}[\kappa V(T, \frac{1}{\kappa} \zeta^*)]$. □

LEMMA A.4. *A random variable $\zeta^*$ is in $\mathcal{D}_{t \to T} \cap \mathbb{L}^1_+$ if and only if there exists a local martingale measure $\mathbb{Q} \in \mathcal{M}^a_T$ and a random variable $\eta \in \mathbb{L}^1_+(\mathcal{F}_t)$ such that*

$$\zeta^* = \eta Z^{\mathbb{Q}}_T / Z^{\mathbb{Q}}_t.$$

PROOF. Suppose, first, that $\zeta^* = \eta Z^{\mathbb{Q}}_T / Z^{\mathbb{Q}}_t$ for some $\eta \in \mathbb{L}^1_+(\mathcal{F}_t)$ and $\mathbb{Q} \in \mathcal{M}^a_T$. In order to show that $\zeta^* \in \mathcal{D}_{t \to T}$, pick a $\rho \in \mathcal{C}_{t \to T}$ of the form $\rho = \int_t^T \pi_u \, dS_u - \zeta$ for some $\pi \in \mathcal{A}_{\mathrm{bd}}$ and $\zeta \in \mathbb{L}^\infty_+$. Then

$$\mathbb{E}[\zeta^* \rho] \leq \mathbb{E}\left[\eta Z^{\mathbb{Q}}_T / Z^{\mathbb{Q}}_t \int_t^T \pi_u \, dS_u\right] = \mathbb{E}\left[\eta \mathbb{E}_{\mathbb{Q}}\left[\int_t^T \pi_u \, dS_u \Big| \mathcal{F}_t\right]\right] = 0,$$

by boundedness of $\int_t^T \pi_u \, dS_u$. Therefore, $\zeta^* \in \mathcal{D}_{t \to T}$.

Conversely, let $\zeta^*$ be an element of $\mathcal{D}_{t \to T} \cap \mathbb{L}^1_+$. We pick an arbitrary $\mathbb{Q}' \in \mathcal{M}^e_T$ and define the random variable $\zeta^*_{\mathbb{Q}'} \in \mathbb{L}^1_+$ by

$$\zeta^*_{\mathbb{Q}'} = \lambda \zeta^*, \qquad \text{where } \lambda = \frac{Z^{\mathbb{Q}'}_t}{\mathbb{E}[\zeta^* | \mathcal{F}_t]} \mathbf{1}_{\{\mathbb{E}[\zeta^* | \mathcal{F}_t] > 0\}} \in \mathbb{L}^0_+(\mathcal{F}_t).$$

We claim that $\zeta^*_{\mathbb{Q}'}$ is the Radon–Nykodim derivative of a local martingale measure. To substantiate this claim, take an arbitrary $\pi \in \mathcal{A}_{\mathrm{bd}}$ and split $\mathbb{E}[\zeta^*_{\mathbb{Q}'} \int_0^T \pi_u \, dS_u]$ into $\mathbb{E}[\zeta^*_{\mathbb{Q}'} \int_0^t \pi_u \, dS_u]$ and $\mathbb{E}[\zeta^*_{\mathbb{Q}'} \int_t^T \pi_u \, dS_u]$. Then

$$\mathbb{E}\left[\zeta^*_{\mathbb{Q}'} \int_0^t \pi_u \, dS_u\right] = \mathbb{E}\left[\mathbb{E}\left[\zeta^*_{\mathbb{Q}'} \int_0^t \pi_u \, dS_u \Big| \mathcal{F}_t\right]\right]$$

$$(A.7) \qquad = \mathbb{E}\left[\mathbb{E}\left[\frac{d\mathbb{Q}'}{d(\mathbb{P}|_{\mathcal{F}_t})} \Big| \mathcal{F}_t\right] \int_0^t \pi_u \, dS_u\right]$$

$$= \mathbb{E}_{\mathbb{Q}'}\left[\int_0^t \pi_u \, dS_u\right] = 0.$$

For the second summand, we define the process $\{\hat{\pi}_u\}_{u \in [0, \infty)}$ by $\hat{\pi}_u = \pi_u \lambda \mathbf{1}_{(t, \infty)}(u)$, for $u \geq 0$. Then $\hat{\pi}$ is predictable and $S$-integrable, and

$$\lambda \int_t^T \pi_u \, dS_u = \int_t^T \hat{\pi}_u \, dS_u.$$

Similarly, processes $\hat{\pi}^n$, defined by $\hat{\pi}^n_u = \pi_u \lambda \mathbf{1}_{\{-n \leq \lambda \leq n\}} \mathbf{1}_{(t, \infty)}(u)$, for $u \geq 0$, are also predictable and $S$-integrable. While the same cannot be concluded for $\hat{\pi}$, all $\hat{\pi}^n$ are in $\mathcal{A}_{\mathrm{bd}}$. So, $\mathbb{E}[\zeta^* \int_t^T \hat{\pi}^n_u \, dS_u] = 0$ and $\int_t^T \hat{\pi}_u \, dS_u = \lim_n \int_t^T \hat{\pi}^n_u \, dS_u$, a.s. Hence,

$$\mathbb{E}\left[\zeta^*_{\mathbb{Q}'} \int_t^T \pi_u \, dS_u\right] = \mathbb{E}\left[\zeta^* \int_t^T \hat{\pi}_u \, dS_u\right] = \mathbb{E}\left[\lim_n \zeta^* \int_t^T \hat{\pi}^n_u \, dS_u\right]$$

$$(A.8) \qquad = \lim_n \mathbb{E}\left[\zeta^* \int_t^T \hat{\pi}^n_u \, dS_u\right] = 0.$$

The above interchange of the limit and the expectation operator is due to the dominated convergence theorem which can be used because

$$\left|\zeta^* \int_t^T \hat{\pi}_u^n \, dS_u\right| \leq \left|\zeta^* \int_t^T \hat{\pi}_u \, dS_u\right| = \zeta^* \lambda \left|\int_t^T \pi_u \, dS_u\right|$$

$$\leq \zeta_{\mathbb{Q}'}^* \left\|\int_t^T \pi_u \, dS_u\right\|_{\mathbb{L}^\infty} \in \mathbb{L}^1.$$

We combine equations (A.7) and (A.8) to obtain $\mathbb{E}[\zeta_{\mathbb{Q}'}^* \int_0^T \pi_u \, dS_u] = 0$, for all $\pi \in \mathcal{A}_{\mathrm{bd}}$. A standard localization argument can be employed to conclude that each component of $S$ is a $\mathbb{Q}$-local martingale, where $\frac{d\mathbb{Q}}{d(\mathbb{P}|_{\mathcal{F}_t})} = \frac{\zeta_{\mathbb{Q}'}^*}{\mathbb{E}[\zeta_{\mathbb{Q}'}^*]}$. Thus,

$$\zeta^* = \frac{1}{\lambda}\zeta_{\mathbb{Q}'}^* = \eta Z_T^{\mathbb{Q}}/Z_t^{\mathbb{Q}}, \qquad \text{where } \eta = \frac{1}{\lambda}Z_t^{\mathbb{Q}}.$$

Finally, $\eta \in \mathbb{L}^1$ since $\zeta^*, Z_T^{\mathbb{Q}}/Z_t^{\mathbb{Q}} \in \mathbb{L}^1$ and $\mathbb{E}[Z_T^{\mathbb{Q}}/Z_t^{\mathbb{Q}}|\mathcal{F}_t] = 1$, a.s. $\square$

THEOREM A.5. *The following relationship holds for the value functions $u$ and $v$ for all $\xi \in \mathbb{L}^\infty(\mathcal{F}_t)$:*

(A.9) $$u(\xi;t,T) = \operatorname*{ess\,inf}_{\eta \in \mathbb{L}_+^1(\mathcal{F}_t)} (v(\eta;t,T) + \xi\eta) \qquad a.s.$$

*Moreover, for each $\xi \in \mathbb{L}^\infty(\mathcal{F}_t)$ there exist $\hat{\eta} \in \mathbb{L}_+^1(\mathcal{F}_t)$ with $\{\hat{\eta} = 0\} \supseteq \{u(\xi;t,T) = \infty\}$ and $\hat{\mathbb{Q}} \in \mathcal{M}_T^a$ such that*

$$u(\xi;t,T) = \mathbb{E}[V(\hat{\eta}Z_T^{\mathbb{Q}}/Z_t^{\mathbb{Q}})|\mathcal{F}_t] + \xi\hat{\eta} = v(\hat{\eta};t,T) + \xi\hat{\eta}.$$

PROOF. We first establish the equality in (A.9). The relationship $U(T,x) \leq V(T,y) + xy$ holds for the functions $U(T,\cdot)$ and $V(T,\cdot)$ for all $x \in \mathbb{R}$, $y \geq 0$, a.s. So, for any $\rho \in \mathcal{C}_{t \to T}$, $\eta \in \mathbb{L}_+^1(\mathcal{F}_t)$ and $\mathbb{Q} \in \mathcal{M}_T^a$,

$$\mathbb{E}[U(T,\xi+\rho)|\mathcal{F}_t] \leq \mathbb{E}[V(T,\eta Z_T^{\mathbb{Q}}/Z_t^{\mathbb{Q}})|\mathcal{F}_t] + \mathbb{E}[(\xi+\rho)\eta Z_T^{\mathbb{Q}}/Z_t^{\mathbb{Q}}|\mathcal{F}_t]$$

$$\leq \mathbb{E}[V(T,\eta Z_T^{\mathbb{Q}}/Z_t^{\mathbb{Q}})|\mathcal{F}_t] + \xi\eta \qquad \text{a.s.}$$

It follows that the left-hand side of (A.9) is at most as large as the right-hand side, a.s. To prove their equality, suppose, contrary to the claim, that there exists an $\mathcal{F}_t$-measurable set $A$ and an $\varepsilon > 0$ such that

(A.10) $$\mathbb{E}[U(T,\xi+\rho)|\mathcal{F}_t] + \varepsilon \mathbf{1}_A < \mathbb{E}[V(T,\eta Z_T^{\mathbb{Q}}/Z_t^{\mathbb{Q}})|\mathcal{F}_t] + \xi\eta \qquad \text{a.s.}$$

for any $\rho \in \mathcal{C}_{t \to T}$, $\eta \in \mathbb{L}_+^1(\mathcal{F}_t)$ and $\mathbb{Q} \in \mathcal{M}_T^a$. The set $A$ has the property that $u(\xi;t,T) < \infty$, a.s., on $A$, and we can assume without loss of generality that there exists $M < \infty$ such that $u(\xi;t,T) \leq M$, a.s., on $A$. After multiplying the inequality $\mathbb{E}[U(T,\xi+\rho)|\mathcal{F}_t] + \varepsilon \mathbf{1}_A < \mathbb{E}[V(T,\eta Z_T^{\mathbb{Q}}/Z_t^{\mathbb{Q}})|\mathcal{F}_t] + \xi\eta$ throughout by $\kappa = \mathbf{1}_A$, noting that $\kappa = 1/\kappa$ on $A$ and taking expectations, we get

$$\mathbb{U}_\kappa(\xi + \rho) + \varepsilon \mathbb{P}[A] < \mathbb{E}\left[\kappa V\left(T, \frac{1}{\kappa}\eta Z_T^\mathbb{Q}/Z_t^\mathbb{Q}\right)\right] + \mathbb{E}[\kappa \eta \xi]$$

for all $\rho \in \mathcal{C}_{t \to T}$, $\mathbb{Q} \in \mathcal{M}_T^a$ and $\eta \in \mathbb{L}_+^1(\mathcal{F}_t)$. This inequality becomes

$$\mathbb{U}_\kappa(\xi + \rho) + \varepsilon \mathbb{P}[A] < \mathbb{V}_\kappa(\zeta^*) + \langle \zeta^*, \xi \rangle,$$

when $\zeta^* = \eta Z_T^\mathbb{Q}/Z_t^\mathbb{Q}$, $\mathbb{Q} \in \mathcal{M}_T^a$ and $\eta \in \mathbb{L}_+^1(\mathcal{F}_t)$ satisfies $\eta = \eta \mathbf{1}_A$. By Proposition A.3 and Lemma A.4, for any other $\zeta^* \in (\mathbb{L}^\infty)^*$ we have $\mathbb{V}_\kappa(\zeta^*) = \infty$. Therefore,

$$u_\kappa(\xi) < u_\kappa(\xi) + \varepsilon \mathbb{P}[A] \leq \inf_{\zeta^* \in (\mathbb{L}^\infty)^*}(\mathbb{V}_\kappa(\zeta^*) + \langle \zeta^*, \xi \rangle).$$

This, however, contradicts Proposition A.1, because $u_\kappa(\xi) \leq M < \infty$.

It remains to justify the second claim of the theorem. Let

$$\kappa = \left(\max\left(1, \operatorname*{ess\,sup}_{\rho \in \mathcal{C}_{t \to T}} \mathbb{E}[U(T, \xi + \rho)|\mathcal{F}_t]\right)\right)^{-1},$$

and let $\hat{\zeta}^*$ be the minimizer of $\zeta^* \mapsto \mathbb{V}_\kappa(\zeta^*) + \langle \zeta^*, \xi \rangle$ over $(\mathbb{L}^\infty)^*$, whose existence follows from Proposition A.1. Since $\mathbb{V}_\kappa(\hat{\zeta}^*) + \langle \hat{\zeta}^*, \xi \rangle = 1$, Proposition A.3 guarantees that $\hat{\zeta}^* \in \mathbb{L}_+^1$. Due to Lemma A.4, there exist $\hat{\mathbb{Q}} \in \mathcal{M}_T^a$ and $\hat{\eta} \in \mathbb{L}_+^1(\mathcal{F}_t)$ with $\{\hat{\eta} = 0\} \supseteq \{\operatorname{ess\,sup}_{\rho \in \mathcal{C}_{t \to T}} \mathbb{E}[U(T, \xi + \rho)|\mathcal{F}_t] = \infty\}$ such that $\zeta^* = \hat{\eta} Z_T^{\hat{\mathbb{Q}}}/Z_t^{\hat{\mathbb{Q}}}$. In order to show that $\hat{\zeta}^*$ attains the essential infimum on the right-hand side of (A.9), assume, to the contrary, that there exist $\varepsilon > 0$, $\eta' \in \mathbb{L}_+^1(\mathcal{F}_t)$, $\mathbb{Q}' \in \mathcal{M}_T^a$ and a set $B \subseteq \mathcal{F}_t$ with $\mathbb{P}[B] > 0$ such that

$$\mathbb{E}[V(T, \eta' Z_T^{\mathbb{Q}'}/Z_t^{\mathbb{Q}'})|\mathcal{F}_t] + \xi \eta' + \varepsilon < \mathbb{E}[V(T, \hat{\eta} Z_T^{\hat{\mathbb{Q}}}/Z_t^{\hat{\mathbb{Q}}})|\mathcal{F}_t] + \xi \hat{\eta} \qquad \text{on } B.$$

Let the random variable $\tilde{\zeta}^* \in \mathbb{L}_+^1$ be defined as

$$\tilde{\zeta}^* = \eta' Z_T^{\mathbb{Q}'}/Z_t^{\mathbb{Q}'} \mathbf{1}_B + \hat{\eta} Z_T^{\hat{\mathbb{Q}}}/Z_t^{\hat{\mathbb{Q}}} \mathbf{1}_{B^c}.$$

It can be shown that $\tilde{\zeta}^* \in \mathcal{D}_{t \to T}$ and, thus, itself of the form $\tilde{\zeta}^* = \tilde{\eta} Z_T^{\tilde{\mathbb{Q}}}/Z_t^{\tilde{\mathbb{Q}}}$, for some $\tilde{\mathbb{Q}} \in \mathcal{M}_T^a$ and $\tilde{\eta} \in \mathbb{L}_+^1(\mathcal{F}_t)$. Using the fact that $\mathbb{E}[V(T, \eta Z_T^\mathbb{Q}/Z_t^\mathbb{Q})|\mathcal{F}_t] = \infty$ on $\{\kappa = 0\}$ for all $\eta, \mathbb{Q}$, we conclude that $B \subseteq \{\kappa > 0\}$. So,

$$\mathbb{V}_\kappa(\tilde{\zeta}^*) + \langle \tilde{\zeta}^*, \xi \rangle \leq \mathbb{V}_\kappa(\hat{\zeta}^*) + \langle \hat{\zeta}^*, \xi \rangle - \varepsilon \mathbb{E}[\kappa \mathbf{1}_B] < \mathbb{V}_\kappa(\hat{\zeta}^*) + \langle \hat{\zeta}^*, \xi \rangle,$$

which contradicts the choice of $\hat{\zeta}^*$ as the minimizer. $\square$

COROLLARY A.6. *For all $\eta \in \mathbb{L}_+^1(\mathcal{F}_t)$, we have*

$$v(\eta; t, T) = \operatorname*{ess\,sup}_{\xi \in \mathbb{L}^\infty(\mathcal{F}_t)}(u(\xi; t, T) - \xi \eta) \qquad a.s.$$

PROOF. Note that Theorem A.5, equation (A.9) in particular, implies that

$$v(\eta;t,T) \geq \operatorname*{ess\,sup}_{\xi \in \mathbb{L}^\infty(\mathcal{F}_t)} (u(\xi;t,T) - \xi\eta) \qquad \text{a.s., for all } \eta \in \mathbb{L}^1_+(\mathcal{F}_t).$$

Let us suppose, contrary to the statement, that there exist $\eta \in \mathbb{L}^1_+(\mathcal{F}_t)$, $\varepsilon > 0$ and a nonnull $\mathcal{F}_t$-measurable set $A$ such that

(A.11) $\mathbb{E}[U(T,\xi+\rho)|\mathcal{F}_t] + \varepsilon \mathbf{1}_A < \mathbb{E}[V(T,\eta Z_T^\mathbb{Q}/Z_t^\mathbb{Q})|\mathcal{F}_t] + \xi\eta$ a.s.

for all $\rho \in \mathcal{C}_{t\to T}$, $\mathbb{Q} \in \mathcal{M}_T^a$ and $\xi \in \mathbb{L}^\infty(\mathcal{F}_t)$. Since we can replace $\eta$ by $\eta\mathbf{1}_A$ without violating the validity of (A.11) on $A$, we assume that $\eta = 0$ on $A^c$. We multiply both sides of (A.11) by $\kappa = \mathbf{1}_A$ and use Proposition A.3 and Lemma A.4 to get

$$\sup_{\rho \in \mathcal{C}_{t\to T}} \mathbb{U}_\kappa(\xi+\rho) + \varepsilon \mathbb{P}[A] < \mathbb{V}_\kappa(\zeta^*) + \langle \xi, \eta \rangle$$

for all $\xi \in \mathbb{L}^\infty(\mathcal{F}_t)$ and all $\zeta^* \in \mathcal{D}^\eta_{t \to T}$. Therefore,

$$\sup_{\xi \in \mathbb{L}^\infty(\mathcal{F}_t)} (u_\kappa(\xi) - \langle \xi, \eta \rangle) \leq v_\kappa(\eta) - \varepsilon \mathbb{P}[A] < v_\kappa(\eta),$$

which is in contradiction with Corollary A.2. □

## APPENDIX B: CLOSEDNESS OF A SET OF STOCHASTIC INTEGRALS

We finish the paper with a technical result needed in the treatment of the Itô-process case of Section 5.

PROPOSITION B.1. *Let the process $\tilde{S}$, the set $K$ and the family $\mathcal{A}^K$ be as in (5.14), (5.15) and the paragraph below it. The set*

$$\tilde{\mathcal{S}} = \left\{ \int_0^\cdot \pi_u \, d\tilde{S}_u : \pi \in \mathcal{A}^K \right\},$$

*is closed with respect to the semimartingale topology.*

PROOF. Were it not for the portfolio constraints, the result would follow directly from Mémin's theorem (see Corollary III.4, page 24 in [23]). With constraints, we need to work a bit harder. Let $\{X^n\}_{n \in \mathbb{N}}$ be given as $X^n = \int_0^\cdot \pi_u^n \, d\tilde{S}_u \in \tilde{\mathcal{S}}$, and let $X$ be a semimartingale on $\mathbb{F}$ such that $X^n \to X$ in the semimartingale topology. By Mémin's theorem, $X$ is of the form

$$X_t = \int_0^t \pi_u \, d\tilde{S}_u$$

for some $\tilde{S}$-integrable predictable process $(\pi_t)_{t\in[0,T]}$. Our task is to show that $\pi_t \in K$ $d\mathbb{P} \times dt$-a.e. By Theorem II.3, page 15 of [23], convergence in the semimartingale topology implies convergence in the space $\mathcal{M}^2 \times \mathcal{A}$ of semimartingales, but only through a subsequence and under an equivalent change of measure. More precisely, there exists a probability measure $\hat{\mathbb{P}} \sim \mathbb{P}$ and an increasing sequence $\{n_k\}_{k\in\mathbb{N}}$ of natural numbers such that

$$\mathbb{E}^{\hat{\mathbb{P}}} \int_0^T [(\pi_{n_k}^0(u) - \pi^0(u))^2 + (\pi_{n_k}^2(u) - \pi^2(u))^2 \\ + |(\pi_{n_k}^0(u) - \pi^0(u))\tilde{\theta}_u| + |\pi_{n_k}^1(u) - \pi^1(u)|]\, du \to 0.$$

An extraction of a further subsequence (still labeled $n_k$) and the fact that the measures $\mathbb{P}$ and $\hat{\mathbb{P}}$ are equivalent yield

$$\pi_{n_k}^i \to \pi^i, \qquad d\mathbb{P} \times dt\text{-a.e.}, i=0,1,2,$$

and so, $\pi_t \in K$, $d\mathbb{P} \times dt$-a.e. $\square$

DEPARTMENT OF MATHEMATICS
UNIVERSITY OF TEXAS AT AUSTIN
1 UNIVERSITY STATION, C1200
AUSTIN, TEXAS
USA
E-MAIL: gordanz@math.utexas.edu